\documentclass[12pt]{article}
\input epsfig.sty

\font\grande=cmr9.5 scaled \magstep4
\font\medio=cmr9.5 scaled \magstep2
\outer\def\beginsection#1\par{\medbreak\bigskip
      \message{#1}\leftline{\bf#1}\nobreak\medskip
\vskip-\parskip
      \noindent}
\begin{document}
\bibliographystyle {unsrt}

\titlepage

\begin{flushright}
CERN-PH-TH/2014-054
\end{flushright}

\vspace{10mm}
\begin{center}
{\grande Tensor B mode and stochastic Faraday mixing}\\
\vspace{10mm}
 Massimo Giovannini 
 \footnote{Electronic address: massimo.giovannini@cern.ch} \\
\vspace{0.5cm}
{{\sl Department of Physics, Theory Division, CERN, 1211 Geneva 23, Switzerland }}\\
\vspace{1cm}
{{\sl INFN, Section of Milan-Bicocca, 20126 Milan, Italy}}
\vspace*{1.5cm}

\end{center}

\centerline{\medio  Abstract}
\vspace{5mm}
This paper investigates the Faraday effect  as a different source of B mode polarization.
The E mode polarization is Faraday rotated provided a stochastic large-scale magnetic field is present prior to photon decoupling.
In the first part of the paper we discuss the case where the tensor modes of the geometry are absent and we argue that the  B mode recently detected by the Bicep2 collaboration cannot be explained by a large-scale magnetic field rotating, through the Faraday effect, the well established 
E mode polarization. In this 
case, the observed temperature autocorrelations would be excessively distorted by the magnetic field. 
In the second part of the paper the formation of Faraday rotation is treated as a stationary, random and Markovian process
with the aim of generalizing a set of scaling laws originally derived in the absence of the tensor modes of the geometry.
We show that the scalar, vector and tensor modes of the brightness perturbations can all be Faraday rotated even if the vector and tensor 
parts of the effect have been neglected, so far, by focussing the attention on the scalar 
aspects of the problem. The mixing between the power spectra of the E mode and B mode polarizations involves a unitary 
transformation depending nonlinearly on the Faraday rotation rate. The present approach is suitable for a general scrutiny 
of the polarization observables and of their frequency dependence.

\noindent

\vspace{5mm}

\vfill
\newpage

\renewcommand{\theequation}{1.\arabic{equation}}
\setcounter{equation}{0}
\section{Introduction}
\label{sec1}
The Bicep2 collaboration has recently reported the detection of a B mode polarization 
that has been attributed to the presence of gravitational waves of inflationary origin \cite{BICEP2,BICEP1} (see also
\cite{SPTpol} for the detection of the B mode coming from lensing).
The tensor fluctuations of the geometry are able to generate a B mode polarization
in the Cosmic Microwave Background (CMB in what follows) 
provided the typical wavelengths of the relic gravitational waves were larger than the 
Hubble radius after matter radiation equality but before decoupling, i.e. at the moment 
when the initial conditions of the polarization anisotropies are set. 

In principe the B mode polarization maybe the result of a more mundane process 
well known in the treatment of cold plasmas, namely the Faraday effect. 
According to the Faraday effect the polarization plane of the incoming radiation
is rotated because of the presence of a magnetic field in a medium 
with finite density of charge carriers. The latter requirement 
is met by the predecoupling plasma that is globally neutral 
but intrinsically charged since the electrons and the ions 
have a common concentration that is ${\mathcal O}(10^{-10})$ the concentration 
of the photons at the corresponding epoch. The only further assumption to get a B mode 
polarization is therefore the presence of a magnetic field that will be assumed 
to be stochastic not to conflict with the assumed isotropy of the background geometry.

The Faraday effect of the CMB polarization was analyzed almost two decades ago 
and, to some extent, even before (see \cite{far1} and references therein).
These suggestions  and have been subsequently discussed in a number of 
different articles (see \cite{far2} and references therein). There are, in principle, other sources
of B mode polarization due to the magnetic fields but they are smaller than 
the contribution of the Faraday effect.

The first task of the present paper will therefore be to establish if the observed B mode polarization
can be ascribed to the Faraday effect. The answer to the question will be, in short, that the 
observed B mode polarization cannot be attributed predominantly to a Faraday rotated 
E mode polarization since the magnetized temperature autocorrelations would be 
too distorted. 

In the second and more technical part of the paper the attention will then be focussed on the 
interference of the Faraday effect  with the B mode produced by the tensor 
mode of the geometry. The aim will be to derive a set of scaling rules that could be 
directly applied to the E mode and B mode power spectra. The Faraday rotation will be described terms of
a stationary, quasi-Markovian and random process \cite{stochde1}.
It will be shown that the evolution of the brightness perturbations obeys a set of stochastic differential equations that can be 
solved using the cumulant expansion \cite{stochde2,stochde3}, 
pioneered in similar contexts by Van Kampen. 

The stochastic approach to the Faraday  effect has been 
exploited in astrophysics where the source of linear 
polarization is provided by the properties synchrotron emission \cite{SYNC1,SYNC2,SYNC3}.
It has been recently suggested \cite{RC1}  that a consistent stochastic description can be successfully 
achieved in the case of the Cosmic Microwave Background (CMB in what follows) where the linear 
polarization is primarily provided by the adiabatic initial conditions of the Einstein-Boltzmann hierarchy \cite{WMAP9}.

The B mode induced by the stochastic Faraday effect thanks to the presence 
of the linear polarization of the CMB can be expressed, according to the results 
of \cite{RC1} as:
\begin{equation}
C_{\ell}^{(EE)} = e^{- \omega_{F} }\, \cosh{\omega_{F}} \,\overline{C}_{\ell}^{(EE)}, \qquad
C_{\ell}^{(BB)} = e^{- \omega_{F} }\, \sinh{\omega_{F}} \,\overline{C}_{\ell}^{(EE)},
\label{INT1}
\end{equation}
where $\overline{C}_{\ell}^{(EE)}$ denotes the autocorrelation of the E mode polarization obtained in the 
absence of stochastic Faraday term and $\omega_{F}$ is given by:
\begin{equation}
 \omega_{F} = 4 \int_{\tau_{r}}^{\tau} \, d\tau_{1} \,  \int_{\tau_{r}}^{\tau} \, d\tau_{2} \langle X_{F}(\tau_{1}) \, X_{F}(\tau_{2}) \rangle.
 \label{INT2}
 \end{equation}
In Eq. (\ref{INT2}) $X_{F}(\tau)$ denotes the Faraday rotation rate and the stochastic process has been assumed, for sake of simplicity, homogeneous 
in space. The derivation of Eq. (\ref{INT1}) 
does not demand  $\omega_{F}\ll 1$ and shows that the stochastic Faraday rate affects not only 
the B mode polarization but, to some extent, also the E mode itself. 
Equations (\ref{INT1}) and (\ref{INT2}) have been derived in \cite{RC1} by assuming that the sole sources of linear 
polarization were the scalar fluctuations of the geometry. It was anticipated in \cite{RC1} that 
the results could be extended to the case where the  initial source of polarization is not only provided 
by the scalar modes but also by the tensor modes that appear in one of the minimal 
extensions of the so-called $\Lambda$CDM paradigm, where $\Lambda$ stands for the dark energy 
component and CDM for the cold dark matter contribution. 

While this analysis was in progress there have been claims of detection 
of a primordial B mode polarization by the Bicep2 collaboration \cite{BICEP2} (see also \cite{BICEP1})
complementing the results of the B mode from lensing \cite{SPTpol}. Although these data are in tension with 
other data sets for various reasons,  it seems timely to present a derivation of the analog of Eq. (\ref{INT1}) 
when the sources of polarization is not only provided by the standard adiabatic mode but 
also by the tensor fluctuations of the geometry. The main result of the present analysis can be summarized by 
writing the analog of Eq. (\ref{INT1}) in this extended set-up where the E modes and the B mode power spectra of the 
tensors are included:
\begin{eqnarray}
C_{\ell}^{(EE)} &=& e^{-\omega_{F}} \cosh{\omega_{F}} \, \biggl( \overline{C}_{\ell}^{(EE)} + {\mathcal C}_{\ell}^{(EE)} \biggr) + e^{- \omega_{F}}\, \sinh{\omega_{F}}\,
 {\mathcal C}_{\ell}^{(BB)},
 \nonumber\\
 {\mathcal C}_{\ell}^{(BB)} &=& e^{-\omega_{F}} \sinh{\omega_{F}} \biggl( \overline{C}_{\ell}^{(EE)} + {\mathcal C}_{\ell}^{(EE)} \biggr) + e^{-\omega_{F}} \, \cosh{\omega_{F}}
\, {\mathcal C}_{\ell}^{(BB)};
\label{INT3}
\end{eqnarray}
as in Eq. (\ref{INT1}) $\overline{C}_{\ell}^{(EE)}$ denotes the E mode power spectrum coming from the scalar modes of the geometry while
${\mathcal C}_{\ell}^{(BB)}$ and ${\mathcal C}_{\ell}^{(EE)}$ (both in calligraphic style) denote, respectively, the polarization observables induced 
by the tensor modes of the geometry. Following the standard terminology, the B-mode autocorrelations are denoted by BB. With similar logic, we talk about the TT, TE and EE angular power spectra denoting, respectively, the autocorrelations of the temperature, the autocorrelations of the E mode and their mutual cross correlations. It is appropriate to recall that the tensor modes of the geometry not only produce BB correlations but also EE and TT power spectra (see e.g. \cite{apprais}).
Comparing  Eqs. (\ref{INT2}) and (\ref{INT3}) in the limit $\omega_{F} \to 0$ we can appreciate that the  B mode polarization disappears from Eq. (\ref{INT2}) 
while it persists in Eq. (\ref{INT3}) and it is solely given by the tensor B mode. 

According to Eqs. (\ref{INT2}) and (\ref{INT3}) both the E mode and the B mode polarization 
are frequency dependent since $\omega_{F}$ is proportional to the square of the 
rate and, ultimately, to the fourth power of the comoving wavelength. The stochastic approach 
to the Faraday rate represents an ideal framework for deriving a set of scaling laws only involving 
the measured polarization power spectra.  The present findings support the consistency 
of the whole description and are suitable for a discussion of the Faraday effect when the predominant 
source of the B mode polarization is provided by relic gravitons with wavelengths comparable with the current Hubble 
radius. 

The present paper is organized as follows. In section \ref{sec2} we shall examine, in the light of the Bicep2 findings, 
the Faraday interpretation of the B mode polarization. In section \ref{sec3} we shall corroborate 
the analysis with a numerical discussion. In section \ref{sec4} we shall describe the phenomenon 
of stochastic Faraday mixing. The polarization observables and their scaling properties  will be deduced in section \ref{sec5} while the concluding 
remarks will be collected in section \ref{sec6}.  To avoid digressions 
some relevant technical aspects of the discussions have been collected in the appendices \ref{APPA}, \ref{APPB} and \ref{APPC}.

\renewcommand{\theequation}{2.\arabic{equation}}
\setcounter{equation}{0}
\section{Bicep2 observations and Faraday effect}
\label{sec2}

The normalization provided by the  Bicep2 results \cite{BICEP2} should be satisfied by any plausible 
physical explanation of the observed B mode autocorrelation. The BB power spectrum should be at most ${\mathcal O}(10^{-2}) \,\mu\mathrm{K}^2$ for typical angular scales of the degree. More specifically\footnote{The normalization of the B mode autocorrelation is often reduced to equivalent 
 tensor to scalar ratio $r_{T}$. We prefer to quote here the B mode autocorrelation in its physical units (i.e. $\mu\mathrm{K}^2$)
since we are going to investigate here a complementary perspective of the problem not relying necessarily on the tensor modes of the geometry.} we can estimate that
 \begin{equation}
{\mathcal G}_{B\ell} = \frac{\ell (\ell+1)}{2\pi} C_{\ell}^{(BB)} \simeq (5.07 \pm 1.13)\times 10^{-2}\, \,\mu\mathrm{K}^2, \qquad \ell \simeq 248,
\label{GBB1}
\end{equation}
where $\ell\simeq 248$ denotes the central value of a bunch of multipoles ranging from $231$ to $265$.
For comparison with the observational data, we shall refer, when needed, to the order of magnitude suggested by Eq. (\ref{GBB1}). Indeed,
for the lowest bunch of multipoles (centered around $\ell \simeq 45$) ${\mathcal G}_{B\ell} \simeq 8.6 \times 10^{-3}\, \mu\mathrm{K}^2$. The remaining 
eight bunches of multipoles give values between $1.3\times 10^{-2} \, \mu\mathrm{K}^2$ (for $\ell \simeq 73$) and $3.2 \times 10^{-2} \, \mu\mathrm{K}^2$ (for $\ell\simeq 317$). The observational frequency of Bicep2 is ${\mathcal O}(150)$ GHz
and it is larger or even much larger than the frequencies of some of the previous polarization experiments (see \cite{dasi,cbi,maxip,boom}).   

The maximalist perspective stipulates that  the observed BB correlation is just the result of the Faraday rotated
EE spectrum whose origin stems from the adiabatic (scalar) fluctuations of the geometry. Conversely, in  the minimalist perspective the observed BB 
spectrum is originated by a primordial tensor mode with tensor to scalar ratio $r_{T} \simeq 0.2$: according to this school of thought magnetic fields as 
well as plasma effects are absent.
There is finally a third logical possibility namely that the BB and the EE correlations are mixed by the Faraday effect, as suggested by Eq. (\ref{INT3}) and by 
the viewpoints conveyed in this investigation. The minimalist perspective cannot be excluded, a priori, but it is fair to say that the Bicep2 data are 
in a certain tension with other CMB data set. We shall therefore focus hereunder on the 
remaining two options giving specific criteria for their empirical exclusion. 

In the maximalist perspective the tensor E and B modes disappear from Eq. (\ref{INT3}) (i.e. ${\mathcal C}_{\ell}^{(EE)} \to 0$ and ${\mathcal C}_{\ell}^{(BB)} \to 0$ in Eq. (\ref{INT3})). Since the tensors are totally absent we could use as well Eq. (\ref{INT1}).
The observed B mode 
depends then on $\omega_{F}$ whose explicit value can be roughly estimated as:
\begin{equation}
\omega_{F} \simeq 4 \biggl(\frac{X_{F}(\vec{x},\tau)}{\epsilon'}\biggr)^2, \qquad 
\frac{X_{F}(\vec{x},\tau)}{\epsilon'} = 35.53 \,  \biggl( \frac{\vec{B} \cdot \hat{n}}{\mathrm{nG}}\biggr)  \biggl( \frac{\mathrm{GHz}}{\overline{\nu}}\biggr)^2, 
\label{rot2}
\end{equation}
showing that the actual value of $X_{F}/\epsilon'$ is not necessarily much smaller than $1$ and it is ${\mathcal O}(1)$ for comoving field strengths of a few nG (i.e. $1\, \mathrm{nG} = 10^{-9}\, \mathrm{G}$) and frequencies ${\mathcal O}(10)$ GHz \cite{RC1}. 
Assuming now that $\overline{{\mathcal G}}_{E\ell}$ is well estimated by the measured E mode autocorrelation \cite{WMAP9,quad} we shall have
\begin{equation}
\overline{{\mathcal G}}_{E\ell} =  \frac{\ell (\ell+1)}{2\pi} \overline{C}_{\ell}^{(EE)} \simeq 50\,\mu\mathrm{K}^2, \qquad \mathrm{for} \,\, \ell \simeq \ell_{\mathrm{max}} = 1000.
\label{rot2a}
\end{equation}
The actual value of the maximum of $\overline{{\mathcal G}}_{E\ell}$ is slightly smaller than $50\,\, \mu\mathrm{K}^2$ (i.e. ${\mathcal O}(43)\,\, \mu\mathrm{K}^2$ \cite{WMAP9,quad}).
The estimate (\ref{rot2a}) is purposely generous with the aim of establishing if and how the benchmark value given by Eq. (\ref{GBB1}) can be roughly reproduced 
without any contribution coming from the tensor modes. 
Consequently, Eqs. (\ref{rot2}) and (\ref{rot2a}) imply that the order of magnitude of the BB correlation can be
estimated at the Bicep2 frequency as
\begin{equation}
{\mathcal G}_{B\ell} = e^{- \omega_{F}} \, \sinh{\omega_{F}} \, \overline{{\mathcal G}}_{E\ell}
\simeq 4.9 \times 10^{-4} \times \biggl(\frac{\vec{B}\cdot\hat{n}}{\mathrm{nG}}\biggr)^2 \biggl(\frac{150\, \mathrm{GHz}}{\overline{\nu}}\biggr)^{4} \,\,
\mu\mathrm{K}^2.
\label{rot3}
\end{equation}
The value $50\,\mu\mathrm{K}^2$ used in Eq. (\ref{rot3}) maximizes $\overline{{\mathcal G}}_{E\ell}$ but it corresponds to relatively high multipole 
moments, typically $\ell \sim 10^{3}$; for smaller multipoles (compatible with the Bicep2 observations) we have that $\overline{{\mathcal G}}_{E\ell}< 5 \,\mu \mathrm{K}^2 $. Moreover  the value of $1$ nG for the magnetic field intensity is barely compatible with the distortions produced by a large-scale magnetic field on the 
temperature autocorrelation: we must bear in mind that the scalar inhomogeneities induced by an inhomogeneous magnetic field are the leading source of distortion of the TT, EE, and TE angular power spectra in comparison with vector and tensor modes. The magnetized CMB observables  have been derived for the magnetized adiabatic mode and compared with the available experimental data with the aim of pinning down the properties of the magnetic field. 
In the second and third papers of Ref. \cite{nn} this analysis has been performed, for the first time, using the WMAP five-year data and later confirmed by subsequent analyses and different data sets (see, e.g. \cite{nn1}). 
 
The estimates of Eqs. (\ref{rot2}), (\ref{rot2a}) and (\ref{rot3}) can be further refined. Assuming that the Faraday rate is perturbative, the angular power spectrum of the Faraday rotated E mode can be computed. This approximation boils down to a sharp separation between the moment 
of formation of the polarization from the moment of the Faraday rotation of the produced polarization (see e.g. last two papers of Ref. \cite{far2}).
The result of the discussion is conceptually similar to the one of Eq. (\ref{rot3}) 
but mathematically more accurate as far as the scaling with the multipoles is concerned.  Defining the normalized form of the Faraday rotation rate  ${\mathcal X}_{F}(\hat{n})$, we have that 
\begin{equation}
\langle {\mathcal X}_{F}(\hat{n}_{1}) {\mathcal X}_{F}(\hat{n}_{2})\rangle = \frac{1}{4\pi}\sum_{\ell} (2 \ell + 1)\,C_{\ell}^{(FF)} \,P_{\ell}(\hat{n}_{1} \cdot\hat{n}_{2}), \qquad {\mathcal X}_{F}(\hat{n}) = \frac{ 3}{16 \pi^2 e} \frac{\hat{n} \cdot \vec{B}}{\nu^2},
\label{F14}
\end{equation}
where $C_{\ell}^{(FF)}$ is the angular power spectrum of the normalized rate.
In terms of the power spectrum of the Faraday rate the autocorrelation of the B-mode polarization 
can be computed as 
\begin{equation}
{\mathcal C}_{\ell}^{(BB)} = \sum_{\ell_{1},\,\,\ell_{2}}  {\mathcal Z}(\ell, \ell_1, \ell_{2})\,\,\overline{C}_{\ell_{2}}^{(EE)}\,\, C_{\ell_{1}}^{(FF)},
\label{F14a}
\end{equation}
where ${\mathcal Z}(\ell, \ell_1, \ell_{2})$ contains
also a Clebsch-Gordon coefficient\footnote{For an explicit expression of ${\mathcal Z}(\ell, \ell_1, \ell_{2})$ see, for instance, the discussion contained in Appendix C of the last paper quoted in Ref. \cite{far2}.} while $\overline{C}_{\ell_{2}}^{\mathrm{EE}}$ is the E-mode power spectrum 
already discussed above;  the sum of Eq. (\ref{F14a}) must be conducted in compliance with 
the constraints stemming from the triangle inequality $|\ell_{1} - \ell_{2}| \leq \ell \leq \ell_{1}+ \ell_{2}$.  The Faraday effect can be treated, within this approach, either with uniform magnetic fields or with stochastic magnetic fields and different analyses have been performed starting with the ones of Ref. \cite{far1} (see also \cite{far2} for an incomplete list of references).

Using analytic methods it is possible to estimate Eq. (\ref{F14a}) at small angular scales (i.e.  $\ell_{1}\gg 1$, $\ell_{2} \gg 1$ and $\ell \gg 1$) where 
the Clebsch-Gordon coefficient inside 
 ${\mathcal Z}(\ell, \ell_1, \ell_{2})$ can be evaluated in analogy with the
 semiclassical limit in non relativistic quantum mechanics. This approach to the asymptotics of the Clebsch-Gordon 
coefficients was originally studied by Ponzano and Regge \cite{ponzano} by exploiting 
the connection of the Clebsch-Gordon coefficients with the Wigner $3j$ and $6j$ symbols (see also \cite{gordon}).
This analytical technique  has been exploited in the last paper of Ref. \cite{far2} for the explicit estimates of ${\mathcal Z}(\ell, \ell_1, \ell_{2})$.
The result is 
that the magnetic field must be larger than about $15$ nG for different ranges of the spectral indices if we ought to be compatible with the 
orders of magnitude of  Eq. (\ref{GBB1}). Thus, the semianalytical considerations suggest that to have a chance of observing 
the Bicep2 value of Eq. (\ref{GBB1}) we are led to magnetic field intensities larger than ${\mathcal O}(10)$ nG. 
Let us now corroborate the previous considerations with an explicit numeric evaluation of Eq. (\ref{F14a}).

\renewcommand{\theequation}{3.\arabic{equation}}
\setcounter{equation}{0}
\section{TT correlations and Faraday B mode}
\label{sec3}

\begin{figure}[!ht]
\begin{center}
\begin{tabular}{|c|c|}
      \hline
      \hbox{\epsfxsize = 7.6 cm  \epsffile{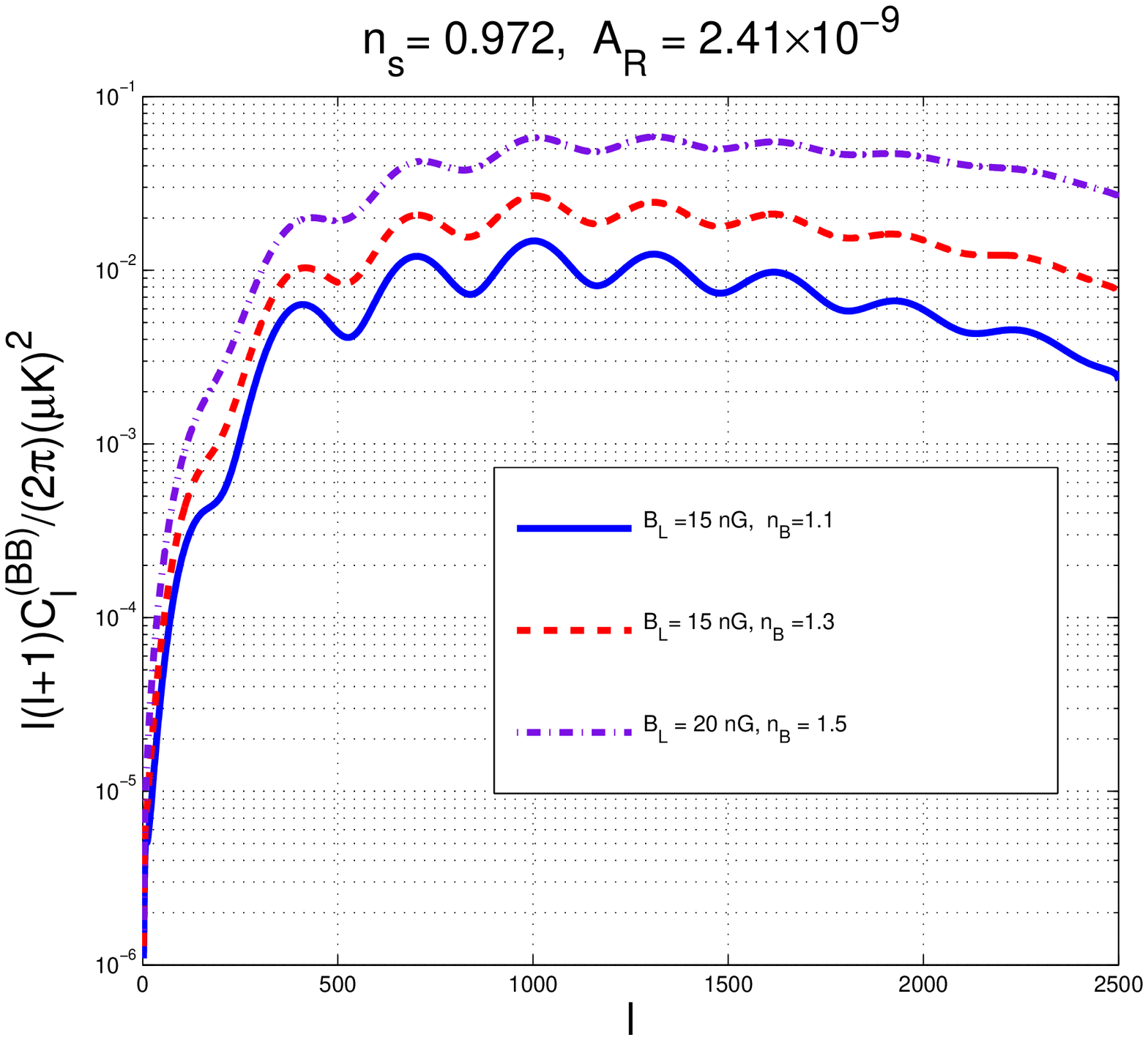}} &
     \hbox{\epsfxsize = 7.6 cm  \epsffile{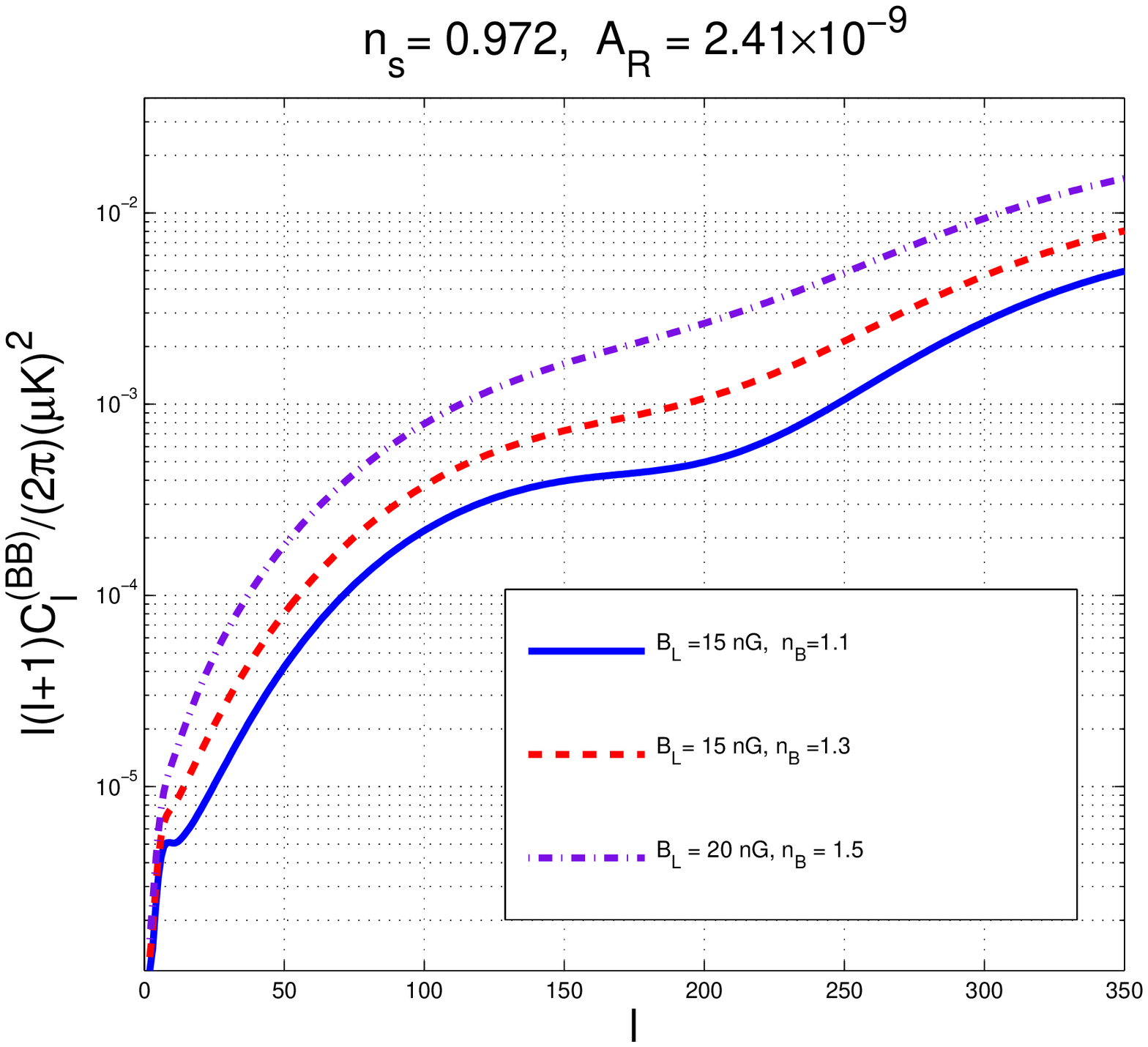}}\\
      \hline
\end{tabular}
\end{center}
\caption[a]{The B mode polarization induced by the Faraday effect  in the magnetized  $\Lambda$CDM scenario with no tensors and with 
different choices of magnetic field parameters.}
\label{Figure1}
\end{figure}
The two-point function of scalar curvature perturbations in Fourier space 
\begin{equation}
\langle {\mathcal R}(\vec{q})\, {\mathcal R}(\vec{k}) \rangle = \frac{2\pi^2}{k^3}\, P_{{\mathcal R}}(k)\,\delta^{(3)}(\vec{q} + \vec{k}),\qquad P_{{\mathcal R}}(k) =  A_{{\mathcal R}} \biggl(\frac{k}{k_{\mathrm{p}}}\biggr)^{n_{\mathrm{s}}-1},
\label{av1a}
\end{equation}
shall be normalized at the pivot scale  $k_{\mathrm{p}}= 0.002\,\, \mathrm{Mpc}^{-1}$;
using the WMAP 9yr data alone \cite{WMAP9} in the light of the concordance scenario we have that $A_{{\mathcal R}}=
(2.41\pm 0.10)\times 10^{-9}$ and $n_{\mathrm{s}} = 0.972\pm0.013$. The exact scale-invariant limit 
is realized when $n_{\mathrm{s}}\to 1$.

The pivotal parameters of the $\Lambda$CDM paradigm can be determined on the basis of different data sets\footnote{ Since we ought to obtain specific quantitive estimates of the B mode polarization in the case of the Bicep2 frequency and for 
the typical parameters of the $\Lambda$CDM paradigm we assume that there are no tensors to begin with. Since 
the Planck collaboration uses anyway the WMAP9 data for the polarization observables we prefer to use directly the WMAP9 data set alone.}
and, for illustrative purposes, we shall consider only three examples. The first one is obtained by 
comparing the $\Lambda$CDM paradigm to the WMAP9 data alone (see, in particular, \cite{WMAP9}):
\begin{equation}
( \Omega_{\mathrm{b}0}, \, \Omega_{\mathrm{c}0}, \Omega_{\mathrm{de}0},\, h_{0},\,n_{\mathrm{s}},\, \epsilon_{\mathrm{re}}) \equiv 
(0.0463,\, 0.233,\, 0.721,\,0.700,\, 0.972,\,0.089),
\label{ppp1}
\end{equation}
with $A_{{\mathcal R}} = 2.41\times 10^{-9}$. If we include the data sets pertaining to the baryon acoustic oscillations (see, e.g. \cite{SDSS}) the string 
of parameters of Eq. (\ref{ppp1})  is slightly different, i.e. $(0.0477,\, 0.247,\, 0.705,\,0.686,\, 0.967,\,0.086)$
with  $A_{{\mathcal R}} = 2.35\times 10^{-9}$.
Another possible set of parameters considered hereunder is the one obtained by combining the WMAP9 data with 
the direct determinations of the Hubble rate leading to 
$(0.0445,\, 0.216,\, 0.740,\,0.717,\, 0.980,\,0.092)$ with  $A_{{\mathcal R}} = 2.45\times 10^{-9}$.  Many other sets of parameters 
corresponding to the combinations of different data sets can be used. These differences are totally immaterial for the present considerations and 
we shall therefore adopt, for illustration, the fiducial set of parameters of Eq. (\ref{ppp1}).

The minimal scenario where the magnetic fields can be consistently included is sometimes dubbed the magnetized $\Lambda$CDM scenario: 
the only two supplementary parameters in comparison with the $\Lambda$CDM parameters are the magnetic 
field amplitude and the magnetic spectral index (see below Eqs. (\ref{av1}) and (\ref{av1a}) for the specific 
definitions). \begin{figure}[!ht]
\begin{center}
\begin{tabular}{|c|c|}
      \hline
      \hbox{\epsfxsize = 7.6 cm  \epsffile{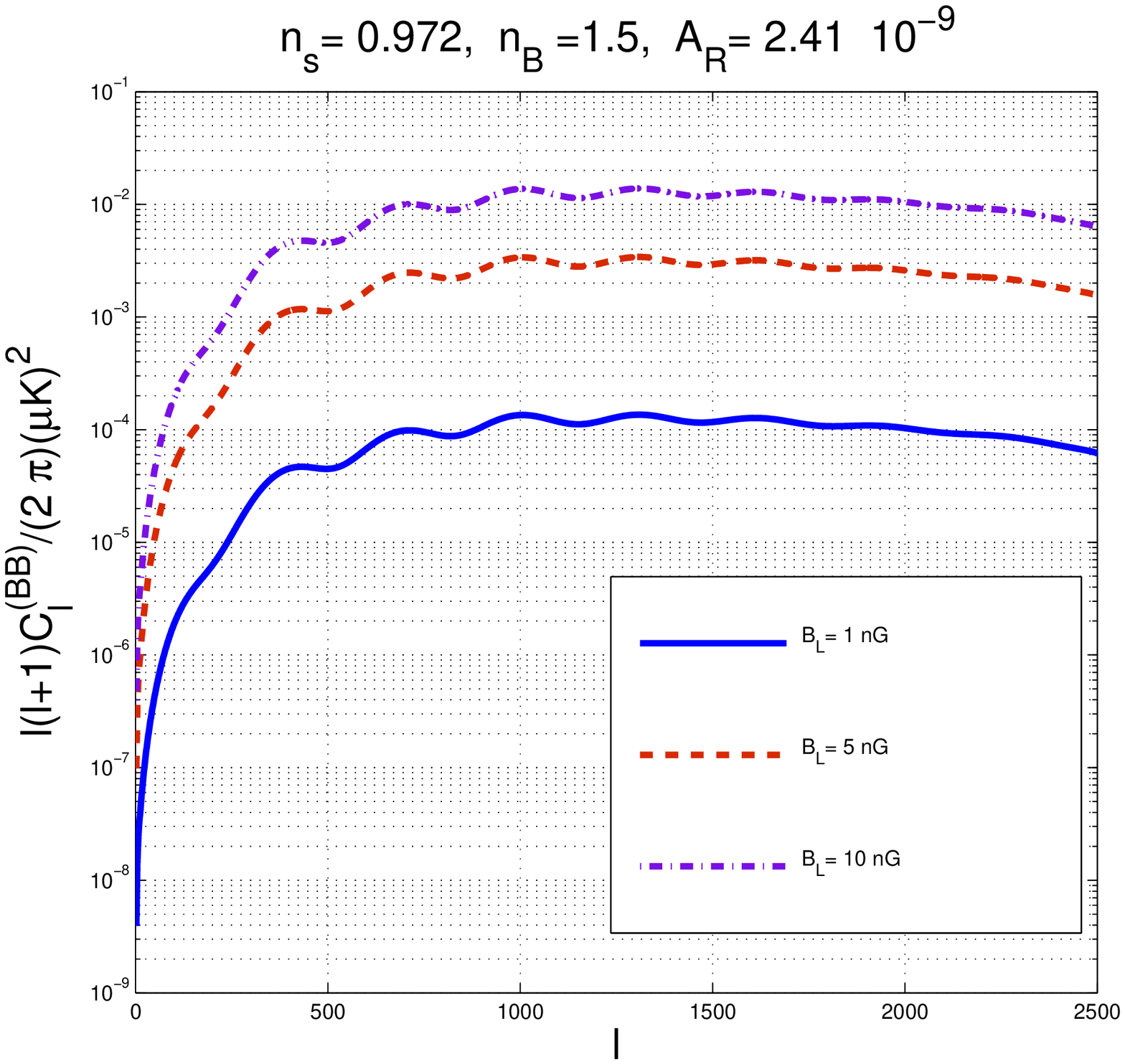}} &
     \hbox{\epsfxsize = 7.6 cm  \epsffile{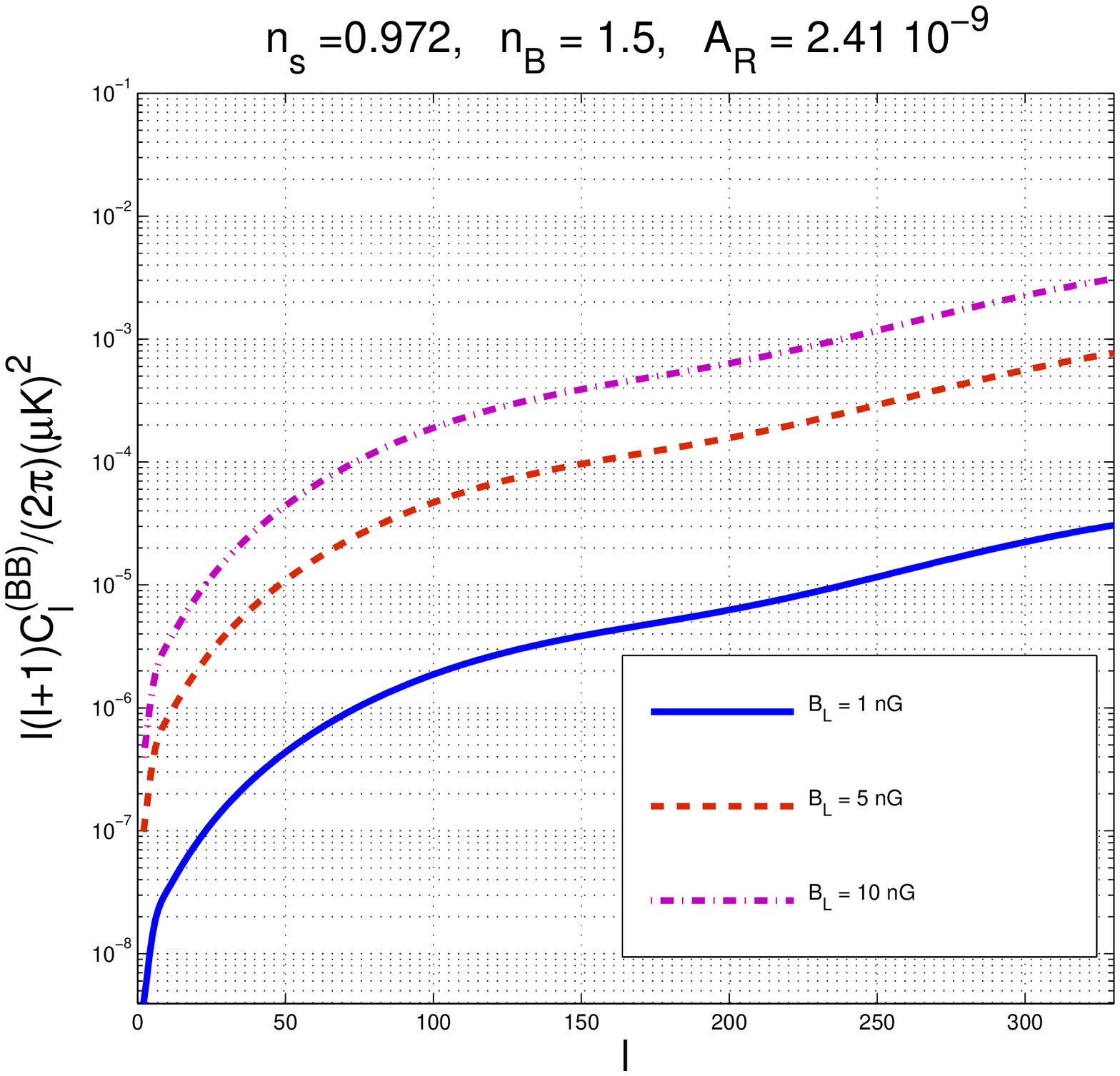}}\\
     \hbox{\epsfxsize = 7.6 cm  \epsffile{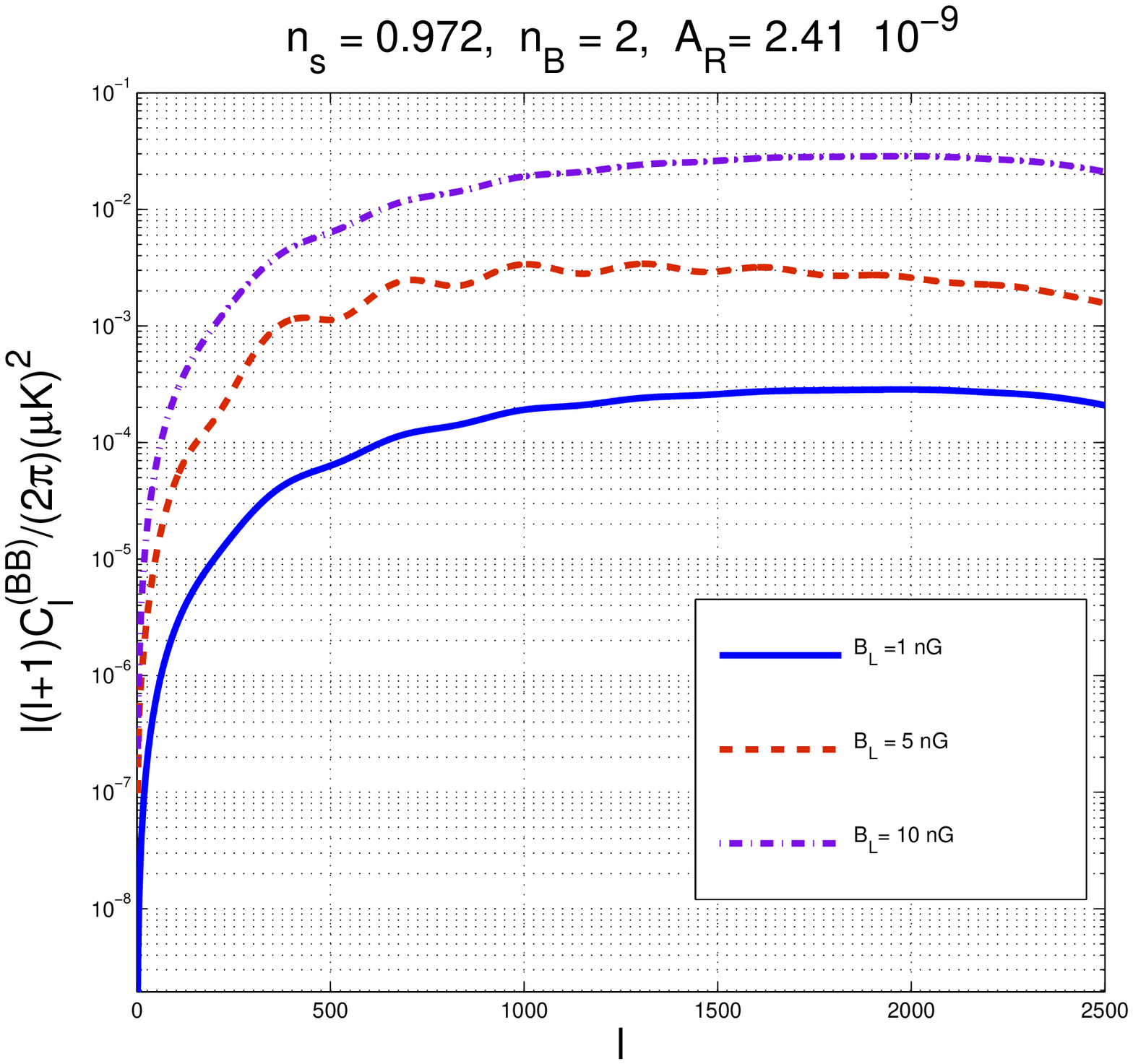}} &
     \hbox{\epsfxsize = 7.6 cm  \epsffile{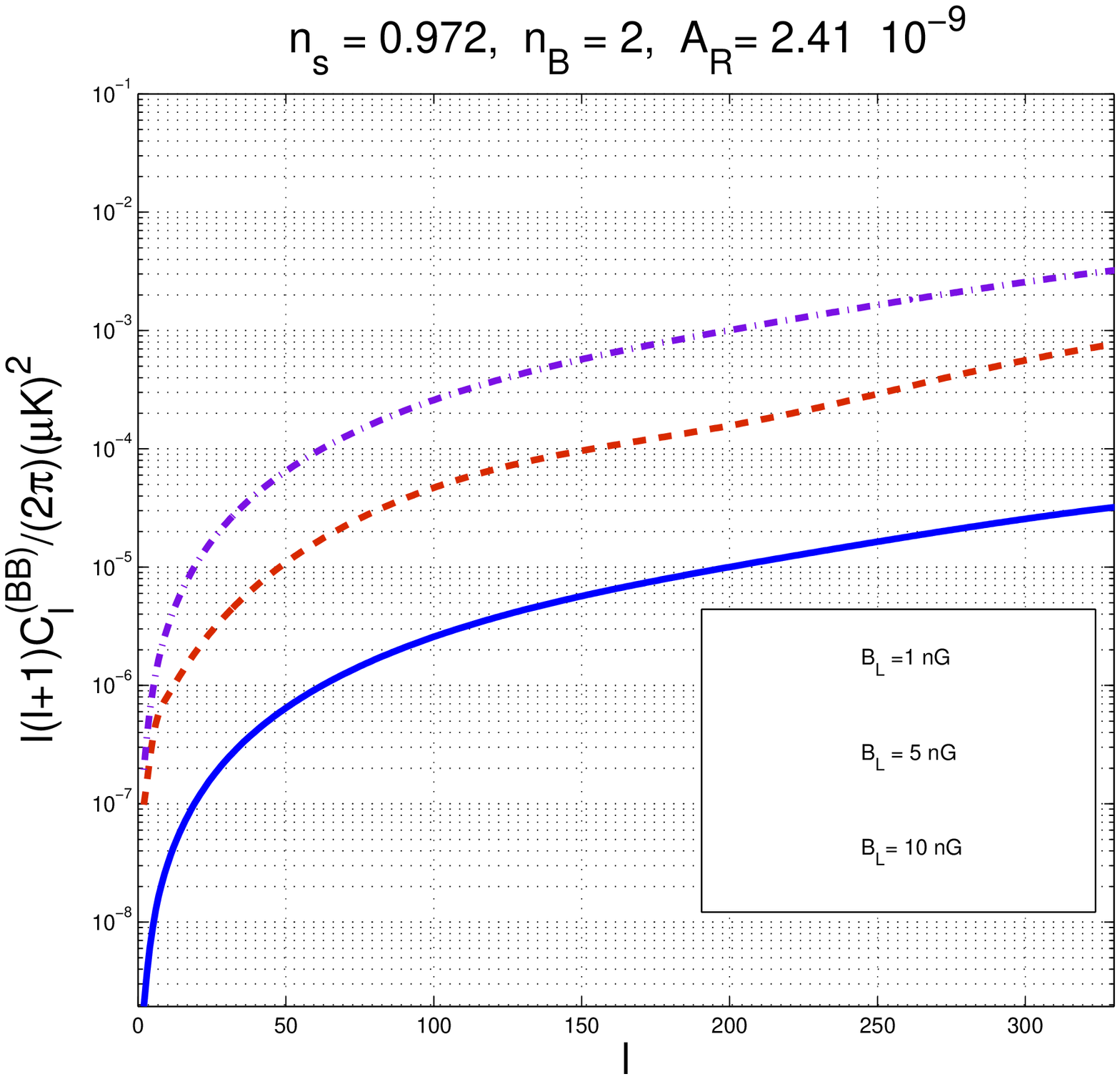}}\\
     \hline
\end{tabular}
\end{center}
\caption[a]{In the two plots at the top the magnetic spectral index is $n_{B} = 1.5$. In the two plots at the bottom $n_{B} =2$. The plots on the left are for small angular scales. The plots 
on the right are for large angular scales. In all the plots the axes are semilogarithmic. }
\label{Figure2}
\end{figure}
The magnetic power spectrum is assigned by following the same conventions of Eq. (\ref{av1a})
\begin{equation}
\langle B_{i}(\vec{q},\tau)\, B_{j}(\vec{k},\tau) \rangle = \frac{2\pi^2}{k^3}\, P_{B}(k,\tau)\, P_{ij}(\hat{k})  \,\delta^{(3)}(\vec{q} + \vec{k}),\qquad 
P_{B}(k,\tau) = A_{B} \biggl(\frac{k}{k_{L}}\biggr)^{n_{B}-1},
\label{av1}
\end{equation}
where $P_{ij}(\hat{k})= ( \delta_{ij} - \hat{k}_{i}\,\hat{k}_{j} )$ and $\hat{k}_{i} = k_{i}/|\vec{k}|$. 
As it can be easily verified from the definition of the Fourier transform, $P_{B}(k,\tau)$ 
has dimensions of an energy density and its square root has, therefore, the dimensions of a field intensity\footnote{ There are some who like to define 
a magnetic power spectrum which is scale invariant for $n\to -3$ and a power spectrum of curvature perturbations which is scale invariant for $n\to 1$.
This will not be the practice followed here: the scale-invariant limit of the power spectra, within the present conventions is $n \to 1$ (see Eqs. (\ref{av1a}) and (\ref{av1})).
These conventions are consistent with the previous literature (see \cite{nn} and references therein).}.
 In Eq. (\ref{av1}), $A_{B}$ has the correct dimensions of an energy density and 
can be related to the regularized magnetic field intensity $B_{\mathrm{L}}$ which is customarily 
employed to phrase the comoving values of the magnetic field intensity.
 In the case when  $n_{\mathrm{B}} > 1$ (i.e. blue magnetic field spectra),  $ A_{B} = 
(2\pi)^{n_{\mathrm{B}} -1} \, B_{\mathrm{L}}^2 /\Gamma[(n_{\mathrm{B}} -1)/2]$; if $n_{\mathrm{B}} < 1$ 
(i.e. red magnetic field spectra), $A_{B} =[ (1 -n_{\mathrm{B}})/2] (k_{\mathrm{p}}/k_{\mathrm{L}})^{(1 - n_{\mathrm{B}})}B_{\mathrm{L}}^2$. In the case of white spectra (i.e. $n_{\mathrm{B}} =1$) the two-point function is logarithmically divergent in real space and this is fully analogous to what happens in Eq. (\ref{av1a}) when $n_{\mathrm{s}} =1$, i.e. the Harrison-Zeldovich (scale-invariant) spectrum.  Quasi-scale invariant spectra with red tilt (i.e. $n_{B} <1$) can arise when magnetic fields are produced in the context of conventional inflationary 
models (see the last paper of Ref. \cite{nn1}) but for numerical purposes we shall mainly focus the attention on blue spectra.

In Fig. \ref{Figure1} with the full, dashed and dot-dashed lines we report the results for the BB spectrum produced by the 
Faraday effect and computed on the basis of Eq. (\ref{F14a}) after having included the magnetic fields in the Einstein-Boltzmann 
hierarchy as in \cite{nn}.
\begin{figure}[!ht]
\begin{center}
      \epsfxsize = 8 cm  \epsffile{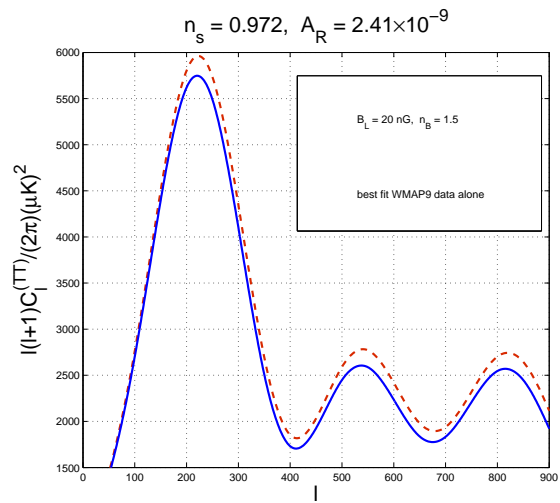}
\end{center}
\caption[a]{The TT correlations of the model illustrated with the dot-dashed line in Fig. \ref{Figure1} is here compared to the best fit to the WMAP9 data alone. }
\label{Figure3}
\end{figure}
In Fig. \ref{Figure1} the values of the magnetic fields range between $15$ and $20$ nG while the magnetic spectral index has been fixed to $n_{B}=1.5$. 
As indicated, the other parameters of Fig. \ref{Figure1} have been fixed to the best fit values given in Eq. (\ref{ppp1}).
Both plots in Fig. \ref{Figure1} share the same parameters but the plot on the right is focussed on the large 
 angular scales while the plot on the left illustrates the small angular scales. Semilogarithmic scales are used in both plots.

Figure \ref{Figure1} suggests that magnetic fields ${\mathcal O}(10)$ nG are unable to reproduce 
the observed Bicep2 amplitude. The analytical estimates of Eqs. (\ref{rot2a})--(\ref{rot3}) and (\ref{F14a}) 
are quantitatively correct but excessively optimistic. 
As anticipated, the EE correlations have been purposely overestimated in Eqs. (\ref{rot2a}) and (\ref{rot3}). 

In Fig. \ref{Figure2} the magnetic spectral index has been fixed at $n_{B}= 1.5$ (plots at the top) and at $n_{B}= 2$ (plots at the bottom). 
The full, dashed and dot-dashed curves in the various plots of Fig. \ref{Figure2} denote, respectively magnetic field intensities of  $1$, $5$ and $10$ nG.
In a frequentistic perspective, beyond the outer contours of the exclusion plots of  the second paper Ref. \cite{nn} the parameters 
of the magnetized $\Lambda$CDM scenario are excluded to 95\% confidence level (see in particular Figs. 1 and 3). For instance
 magnetic field intensities ${\mathcal O}(1.5)$ nG are not excluded only by selecting appropriate values of the spectral indices. Field strengths 
 ${\mathcal O}(10)$ nG and, a fortiori, ${\mathcal O}(20)$ nG are are ruled out in spite of the value of the spectral index. This means that 
if we compute the TT  correlations for the models of Figs. \ref{Figure1} or \ref{Figure2} we shall see that they are excessively distorted. 

According to Fig. \ref{Figure1} the model that is closer to the BICEP2 data is the one illustrated by the dod-dashed line. The TT correlations 
for such a model are compared, in Fig. \ref{Figure3} with the best fit to the WMAP9 data alone. 
The conclusion is that the models of Fig. \ref{Figure1} get close to the Bicep2 measurement but are already excluded by the analysis of the other CMB observables.

Magnetic fields systematically less intense than the ones of Figs. \ref{Figure1} and \ref{Figure2}  lead to B mode polarizations 
that are minute in the Bicep2 region. The value of $B_{L}\simeq {\mathcal O}(\mathrm{nG})$ leads to a B mode signal ${\mathcal O}(10^{-6})\,\,\mu\mathrm{K}^2$. 
 Let us finally check if and how we could find some sort of model that could be barely compatible with the BICEP2 data at 
large scales while still being compatible with the previous bounds on the B mode polarization at smaller scales. 
\begin{figure}[!ht]
\begin{center}
 \epsfxsize = 7.6 cm  \epsffile{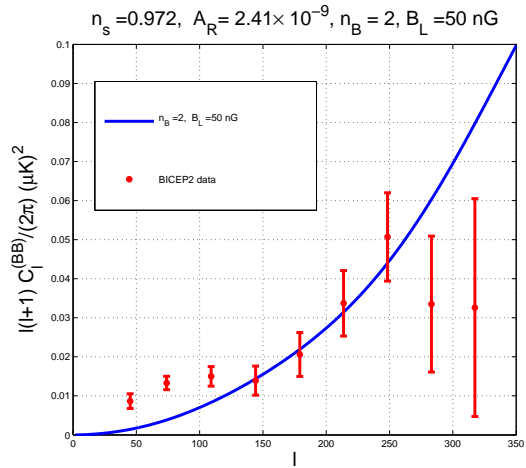}
\end{center}
\caption[a]{The BB correlations for an extreme set of magnetic field parameters.}
\label{Figure4}
\end{figure}
This aspect is discussed in Fig. \ref{Figure4} where the Bicep2 data points seem to be vaguely close to the curve but at smaller angular scales 
scales (i.e. $\ell \sim 10^{3}$) the B mode polarization is ${\mathcal O}(0.8)\,(\mu\mathrm{K})^2$ which is actually an enormous value (potentially 
in conflict with existing upper limits at small angular scales) and leading to ridiculously large distortions 
of the TT correlations of the type illustrated in Fig. \ref{Figure2}.

\renewcommand{\theequation}{4.\arabic{equation}}
\setcounter{equation}{0}
\section{Brightness perturbations}
\label{sec4}
The main objective of the the present section and of the following one will be to derive the results contained in Eq. (\ref{INT3}).
Further details on the cumulant expansion and on the technicalities involved in the calculation are collected in Appendix \ref{APPC}. 
At the end of section \ref{sec5} this technical effort will lead to the derivation of a set of scaling laws that 
can be directly applied to the angular power spectra.

\subsection{General considerations} 
We shall consistently work in a conformally flat space-time whose line element and metric tensor are defined as:
\begin{equation}
ds^2 = g_{\mu\nu} dx^{\mu} dx^{\nu} = a^2(\tau)[ d\tau^2 - d\vec{x}^2],\qquad g_{\mu\nu} = a^2(\tau) \eta_{\mu\nu},
\label{metric}
\end{equation}
$a(\tau)$ is the scale factor of a conformally flat geometry of Friedmann-Robertson-Walker type  and $\tau$ is the conformal time coordinate.
The fluctuations of the Stokes parameters in comparison to their equilibrium values
can be decomposed as: 
\begin{equation}
 \Delta_{X}(\vec{x},\tau) = \Delta^{(\mathrm{s})}_{X}(\vec{x},\tau) + \Delta^{(\mathrm{v})}_{X}(\vec{x},\tau) + \Delta^{(\mathrm{t})}_{X}(\vec{x},\tau),
\label{BRDEC1}
\end{equation}
where $X=I,\,Q,\,U,\,V$ denotes one of the four Stokes parameters and where the superscripts refer, respectively, to the scalar, vector and tensor modes of the geometry. 

The brightness perturbations\footnote{The partial derivatives with respect to $\tau$ will be denoted by 
$\partial_{\tau}$; the partial derivatives with respect to the spatial coordinates 
will be instead denoted by $\partial_{i}$ with $i= 1,\,2,\,3$.}  defined in Eq. (\ref{BRDEC1}) are affected by the presence of the Faraday rotation term $X_{F}$
\begin{eqnarray}
&& \partial_{\tau}\Delta_{I} + (\epsilon' + n^{i} \, \partial_{i} )\Delta_{I} = {\mathcal M}_{I}(\vec{x},\tau)
\label{re1}\\
&& \partial_{\tau} \Delta_{\pm} + (\epsilon' + n^{i} \, \partial_{i} ) \Delta_{\pm} = {\mathcal M}_{\pm}(\vec{x},\tau) \mp 2 i \, X_{F}(\vec{x},\tau) \Delta_{\pm},
\label{re2}\\
&& \partial_{\tau} \Delta_{V} + (\epsilon' + n^{i} \, \partial_{i} ) \Delta_{V} = {\mathcal M}_{V}(\vec{x},\tau),
\label{re3}
\end{eqnarray}
where $\Delta_{\pm}(\vec{x},\tau)  = \Delta_{Q}(\vec{x},\tau) \pm 
i \, \Delta_{U}(\vec{x},\tau)$ and $\epsilon'$ is the differential optical depth
\begin{equation}
\epsilon' = a \tilde{n}_{\mathrm{0}} x_{\mathrm{e}} \sigma_{\gamma\mathrm{e}}, \qquad \sigma_{\gamma\mathrm{e}} = \frac{8}{3} \pi r_{\mathrm{e}}^2, \qquad r_{\mathrm{e}}= \frac{e^2}{m_{\mathrm{e}}}.
\label{diffop}
\end{equation}
In Eq. (\ref{re2}) $X_{F}(\vec{x},\tau)$ denotes the Faraday rotation rate:
\begin{equation}
X_{F}(\vec{x},\tau)= \frac{\overline{\omega}_{Be}}{2} \biggl(\frac{\overline{\omega}_{pe}}{\overline{\omega}}\biggr)^2 =  \frac{e^3}{ 2 \pi} \biggl( 
\frac{n_{e}}{m_{e}^2 \, a^2 }\biggr)\, \biggl( \frac{\vec{B} \cdot \hat{n}}{\overline{\nu}^2}\biggr), 
\label{rot1}
\end{equation} 
where $\overline{\omega}= 2 \pi \overline{\nu}$ is the (comoving) angular frequency, while $\overline{\omega}_{Be}$  and $\overline{\omega}_{pe}$ denote the comoving Larmor and plasma frequencies, respectively; $n_{e} = \tilde{n}_{0} \, a^3$ is the comoving electron concentration.

If the time scale of spatial variation of the rate is comparable with the time scale of spatial variation of the gravitational fluctuations, 
$X_{F}$ can be considered only time dependent (i.e. a stochastic process). In the opposite situation the Faraday rate must be considered fully inhomogeneous (i.e. a stochastic field). Both possibilities will be considered. There is, in principle, also a third case, namely the situation where  
 $X_{F}$  is just a (time-independent) random variable characterized by a given probability distribution. This is, in some sense, the simplest and most naive 
case and it follows from the present results when $X_{F}$ is considered a space-time constant with random distribution. It is finally useful to recall that the stochastic 
Faraday effect has been discussed when $X_{F}$ is a constant random variable in the framework of the synchrotron emission (see, e.g. the second paper of Ref. \cite{SYNC2}).
  
Equations (\ref{re1}), (\ref{re2}) and (\ref{re3}) may also depend on the comoving frequency for a different reason. Since the magnetic field modifies the trajectories of the charged particles it also affects the collision matrix for the photons impinging on the electrons.  The collision matrix will the inherit 
corrections ${\mathcal O}(f_{e}^2)$ where $f_{e}(\overline{\omega})$ is defined as:
\begin{equation}
f_{e}(\overline{\omega}) = \biggl(\frac{\overline{\omega}_{\mathrm{Be}}}{\overline{\omega}}\biggr)
 = 2.79 \times 10^{-12} \biggl(\frac{\hat{n}\cdot \vec{B}}{\mathrm{nG}}\biggr)
 \biggl(\frac{\mathrm{GHz}}{\overline{\nu}}\biggr) (z_{*} +1)\ll 1,
\label{FE}
\end{equation}
and $z_{*}$ is the redshift to last scattering. This effect is well understood but rather difficult to compute when scalar, vector and tensor modes of the geometry are included. Since the typical frequencies 
of the observational channel are normally much larger than the Larmor frequency of the electrons, the corrections ${\mathcal O}(f_{e}^2)$ are negligible. 
This comment anticipates a possible objection and demonstrates that, in this context, the role of the magnetic field in the collision matrix can be ignored at least 
in the first approximation. This analysis has been however performed in a series of papers in connection with the problem of the 
circular polarizations (see \cite{mgsc} and references therein). To be precise the exact form of the correction ${\mathcal O}(f_{e}^2)$ to the full equations will be reported below in the scalar and tensor cases.

\subsection{Scalar brightness perturbations}
The scalar equations can be written, in a compact form, as:
\begin{eqnarray}
&&\partial_{\tau} \Delta^{(\mathrm{s})}_{I} + (\epsilon' + n^{i} \, \partial_{i} ) \Delta^{(\mathrm{s})}_{I} = {\mathcal M}^{(\mathrm{s})}_{I}(\vec{x},\tau),
\label{SC0}\\
&&\partial_{\tau} \Delta^{(\mathrm{s})}_{\pm} + (\epsilon' + n^{i} \, \partial_{i} ) \Delta^{(\mathrm{s})}_{\pm} = {\mathcal M}_{P}^{(\mathrm{s})}(\vec{x},\tau) \mp 2 i \, X_{F}(\vec{x},\tau) \Delta^{(\mathrm{s})}_{\pm}.
\label{SC1}
\end{eqnarray}
The evolution equations of the polarization do not contain explicitly the fluctuations of the metric that appear instead in the evolution of $\Delta^{(\mathrm{s})}_{I}$ and have 
been included in the source term  ${\mathcal M}^{(\mathrm{s})}_{I}(\vec{x},\tau)$ that will play little role in the forthcoming considerations (see, however, Eqs. 
(\ref{BRDEC6}) and (\ref{BRDEC9a})).
The dependence of the polarization observables on the metric fluctuations appears instead indirectly through the multipoles of $\Delta^{(\mathrm{s})}_{I}$ entering $ {\mathcal M}^{(\mathrm{s})}_{\pm}(\vec{x},\tau)$.  

Denoting with $\delta^{(\mathrm{s})}_{I}$ and $\delta^{(\mathrm{s})}_{\pm}$ the Fourier transforms of the corresponding brightness perturbations, 
the collision terms in the scalar case can be computed from Eqs. (\ref{AP2}) and (\ref{AP3}) of appendix \ref{APPA} and they are 
\begin{eqnarray}
&&C_{I}^{(\mathrm{s})} = \frac{3 \epsilon'}{8 \pi} \int d\Omega'\, \sum_{J = I,\, Q,\, U} {\mathcal N}_{I J} {\mathcal I}^{(\mathrm{s})}_{J} =\epsilon'\bigg[ \delta_{I 0}  - \frac{P_{2}(\mu)}{2} S^{(\mathrm{s})}_{\mathrm{P}} + {\mathcal O}(f_{e}^2) \biggr],
\label{CSI}\\
&&C_{Q}^{(\mathrm{s})} = \frac{3 \epsilon'}{8 \pi} \int d\Omega'\, \sum_{J = I,\, Q,\, U} {\mathcal N}_{Q J} {\mathcal I}^{(\mathrm{s})}_{J} = \frac{3 }{4}(1 -\mu^2) \epsilon'\biggl[ S_{P}^{(\mathrm{s})} +   {\mathcal O}(f_{e}^2) \biggr],
\label{CSQ}\\
&& C_{U}^{(\mathrm{s})} = \frac{3 \epsilon'}{8 \pi} \int d\Omega'\, \sum_{J = I,\, Q,\, U} {\mathcal N}_{U J} {\mathcal I}^{(\mathrm{s})}_{J} =0,
\label{CSU}
\end{eqnarray}
where  the corrections ${\mathcal O}(f_{e}^2)$ have been only indicated but will be explicitly included in the full equations (see below); the term 
$S^{(\mathrm{s})}_{P}$ denotes the standard combination:
\begin{equation}
S^{(\mathrm{s})}_{P} = \delta^{(\mathrm{s})}_{I\,2} + \delta^{(\mathrm{s})}_{Q\,0} + \delta^{(\mathrm{s})}_{Q\,2},  
 \label{SsP}
 \end{equation}
The conventions adopted for the definition of the multipoles of the polarization and of the intensity include the factor $(2 \ell+1)$ in the expansion, i.e.
\begin{eqnarray}
&& \delta^{(\mathrm{s})}_{P}(\mu,k,\tau)  = \sum_{\ell} (-i)^{\ell} (2\ell + 1) \, P_{\ell}(\mu)\, \delta^{(\mathrm{s})}_{P\,\ell}(k,\tau),
\label{ZZ1}\\
&& \delta^{(\mathrm{s})}_{I}(\mu,k,\tau)  = \sum_{\ell} (-i)^{\ell} (2\ell + 1) \, P_{\ell}(\mu)\, \delta^{(\mathrm{s})}_{I\,\ell}(k,\tau),
\label{TT1}
\end{eqnarray}
where $P_{\ell}(\mu)$ denote the Legendre polynomials. When $X_{F}(\tau)$  is a homogenous stochastic process the equation for the intensity is\footnote{We have also 
introduced for completeness and in analogy with the forthcoming tensor case the contribution coming from the scalar metric fluctuations 
and modifying the collisionless part of the Boltzmann equation. The corresponding details and the related conventions are collected in appendix 
\ref{APPB}.}
\begin{eqnarray}
\partial_{\tau} \delta^{(\mathrm{s})}_{I}
+ ( i k\mu + \epsilon') \delta^{(\mathrm{s})}_{I} &=& \partial_{\tau} \psi - i k \mu \phi 
\nonumber\\
&+& \epsilon' \biggl[ \delta^{(\mathrm{s})}_{I 0} + \mu v_{\mathrm{b}} - \frac{P_{2}(\mu)}{2} S^{(\mathrm{s})}_{\mathrm{P}} 
+ f_{\mathrm{e}}^2 \biggl(\frac{2}{3} \delta_{I 0} + \frac{P_{2}(\mu)}{6} S^{(\mathrm{s})}_{\mathrm{P}}\biggr)\biggr],
\label{DDI}
\end{eqnarray}
while the equations for the polarization are:
\begin{eqnarray}
&&  \partial_{\tau} \delta^{(\mathrm{s})}_{Q} + ( i k\mu + \epsilon') \delta^{(\mathrm{s})}_{Q} = \frac{(3 - f_{e}^2)}{4}(1 -\mu^2) \epsilon' S_{P}^{(\mathrm{s})} + 2 X_{F}(\tau) \delta^{(\mathrm{s})}_{U},
\nonumber\\
&&  \partial_{\tau} \delta^{(\mathrm{s})}_{U} + ( i k\mu + \epsilon') \delta^{(\mathrm{s})}_{U} = - 2 X_{F}(\tau) \delta^{(\mathrm{s})}_{Q}.
\label{DDU}
\end{eqnarray}
In Eq. (\ref{DDU}) we included also the correction (parametrized by $f_{e}^2$) arising when we take into account the magnetic field 
in the scattering term.

If $X_{F}(\tau)$ is a spatially homogenous stochastic process the relevant equation we shall be dealing with is:
\begin{eqnarray}
&& \partial_{\tau} \delta^{(\mathrm{s})}_{\pm} + (i k \mu + \epsilon') \delta^{(\mathrm{s})}_{\pm} =\frac{(3 - f_{e}^2)}{4}(1 -\mu^2) \epsilon' S_{P}^{(\mathrm{s})}(\vec{k}, \tau) \mp 2 i  \, X_{F}(\tau) \delta^{(\mathrm{s})}_{\pm}.
\label{un2a}
\end{eqnarray}
If $X_{F}$ is a spatially inhomogeneous stochastic process the discussion in mathematically slightly different but physically equivalent as far 
as the frequency scaling is concerned. More specifically, the evolution equations for $\delta_{\pm}$ will now contain a convolution and can be written as:
\begin{equation}
\partial_{\tau}\delta^{(\mathrm{s})}_{\pm} + ( i k \mu +\epsilon') \delta^{(\mathrm{s})}_{\pm} = \frac{3}{4} ( 1 - \mu^2) \, \epsilon'\,S^{(\mathrm{s})}_{P}(\vec{k},\tau) \mp i \, b_{F}(\overline{\nu},\tau) \int d^{3} p \,\delta^{(\mathrm{s})}_{\pm}(\vec{k} + \vec{p},\tau) \, n^{i} B_{i}(\vec{p},\tau),
\label{inhom1}
\end{equation} 
where we introduced, for convenience, the term $b_{F}(\overline{\nu},\tau) =  2 \, e^3 n_{e}/[(2\pi)^{5/2} m_{e}^2 a^2(\tau) \overline{\nu}^2]$ and 
neglected, for simplicity, the corrections ${\mathcal O}(f_{e}^2)$. The addition 
of the spatial dependence is just a technical complication since the essential aspect is the stochastic evolution in time.

\subsection{Tensor brightness perturbations}
The tensor brightness perturbations can still be written in compact notation as:
\begin{eqnarray}
&&\partial_{\tau} \Delta^{(\mathrm{t})}_{I} + (\epsilon' + n^{i} \, \partial_{i} ) \Delta^{(\mathrm{t})}_{I} = {\mathcal M}^{(\mathrm{t})}_{I}(\vec{x},\tau),
\label{T0}\\
&&\partial_{\tau} \Delta^{(\mathrm{t})}_{\pm} + (\epsilon' + n^{i} \, \partial_{i} ) \Delta^{(\mathrm{t})}_{\pm} = {\mathcal M}_{P}^{(\mathrm{t})}(\vec{x},\tau) \mp 2 i \, X_{F}(\vec{x},\tau) \Delta^{(\mathrm{t})}_{\pm}.
\label{T1}
\end{eqnarray}
In the scalar case the evolution of the intensity is coupled to the polalrization only through the 
source term $S^{(\mathrm{s})}_{P}$. In the tensor case something similar happens with the difference 
that the two polarizations of the relic tensor wave will determine also a specific azimuthal dependence 
of the other brightness perturbations.  Recalling the results 
of the appendices A and B we can write, in Fourier space, 
\begin{eqnarray}
&& \partial_{\tau} \delta^{(\mathrm{t})}_{I} + (i k \mu + \epsilon' ) \delta^{(\mathrm{t})}_{I} - \frac{1}{2} \biggl\{ [ (\hat{n}\cdot\hat{a})^2 - (\hat{n}\cdot\hat{b})^2] h_{\oplus}(\vec{k},\tau) + 2 (\hat{n}\cdot\hat{a}) (\hat{n} \cdot \hat{b})
h_{\otimes}(\vec{k},\tau)\biggr\} = C_{I}^{(\mathrm{t})},
\nonumber\\
&& \partial_{\tau} \delta^{(\mathrm{t})}_{Q} + (i k \mu + \epsilon' ) \delta^{(\mathrm{t})}_{Q} = C_{Q}^{(\mathrm{t})} + 2 X_{F} \delta^{(\mathrm{t})}_{U},
\nonumber\\
&& \partial_{\tau} \delta^{(\mathrm{t})}_{U} + (i k \mu + \epsilon' ) \delta^{(\mathrm{t})}_{U} = C_{U}^{(\mathrm{t})} - 2 X_{F} \delta^{(\mathrm{t})}_{Q},
\label{TTU}
\end{eqnarray}
where we have considered the case where $X_{F}(\tau)$ is a stochastic process. 

Before computing the collision terms it is therefore necessary to specify the azimuthal structure of the relevant 
brightness perturbations. Since $(\hat{n}\cdot\hat{a})^2 - (\hat{n}\cdot\hat{b})^2 = (1 - \mu^2) \cos{2 \varphi}$
and $2 (\hat{n}\cdot\hat{a}) (\hat{n} \cdot \hat{b}) = (1- \mu^2) \sin{2 \varphi}$ the intensity brightness 
must be written as
\begin{equation}
\delta_{I}^{(\mathrm{t})}(\vec{k},\tau,\mu,\varphi) = (1 - \mu^2) \biggl[ \cos{2 \varphi} {\mathcal Z}_{\oplus}(k,\tau) + \sin{2 \varphi} {\mathcal Z}_{\otimes}(k,\tau)\biggr].
\label{TTU2}
\end{equation}
The functions ${\mathcal Z}_{\oplus}(k,\tau)$ and ${\mathcal Z}_{\otimes}(k,\tau)$ are obey the same equation since the components 
of the tensor polarization $h_{\oplus}$ and $h_{\otimes}$ obey the same equation. 
Keeping track of all the normalizations we can therefore write Eq. (\ref{TTU2}) as: 
\begin{eqnarray}
\delta_{I}^{(\mathrm{t})}(\vec{k},\tau,\hat{n}) &=& {\mathcal D}_{T}(\hat{n}) \,\delta_{T}^{(\mathrm{t})}(k,\tau),
\label{TTU2a}\\
\delta_{+}^{(\mathrm{t})}(\vec{k},\tau,\hat{n}) &=& {\mathcal D}_{+}(\hat{n}) \,\overline{\delta}_{+}^{(\mathrm{t})}(k,\tau),
\label{TTU2b}\\
\delta_{-}^{(\mathrm{t})}(\vec{k},\tau,\hat{n}) &=& {\mathcal D}_{-}(\hat{n}) \,\overline{\delta}_{-}^{(\mathrm{t})}(k,\tau),
\label{TTU2c}
\end{eqnarray}
where ${\mathcal D}_{T}(\hat{k},\hat{n})$ and ${\mathcal D}_{\pm}(\hat{n})$ are defined as:
\begin{eqnarray}
{\mathcal D}_{T}(\hat{n}) &=& (1 - \mu^2) \biggl[ e^{ 2 i \varphi} {\mathcal F}_{1}(\vec{k}) + e^{- 2 i \varphi} {\mathcal F}_{2}(\vec{k}) \biggr], 
\nonumber\\
{\mathcal D}_{+}(\hat{n}) &=&\biggl[ (1 - \mu)^2 e^{ 2 i \varphi} {\mathcal F}_{1}(\vec{k}) +(1 + \mu)^2  e^{- 2 i \varphi} {\mathcal F}_{2}(\vec{k}) \biggr],
\nonumber\\
{\mathcal D}_{-}(\hat{n}) &=&  \biggl[ (1 + \mu)^2 e^{ 2 i \varphi} {\mathcal F}_{1}(\vec{k}) +(1 - \mu)^2  e^{- 2 i \varphi} {\mathcal F}_{2}(\vec{k}) \biggr].
\label{ang}
\end{eqnarray}
In Eq. (\ref{ang}) two time-independent stochastic variables ${\mathcal F}_{1}(\vec{k}) = ( {\mathcal F}_{\oplus} - i {\mathcal F}_{\otimes})/\sqrt{2}$  and ${\mathcal F}_{2}(k) ( {\mathcal F}_{\oplus} - i {\mathcal F}_{\otimes})/\sqrt{2}$ have been introduced. Each of the two stochastic polarizations ${\mathcal F}_{\oplus}$ and ${\mathcal F}_{\otimes}$ is related to the tensor power spectrum ${\mathcal P}_{T}(k)$ as
\begin{equation}
\langle {\mathcal F}_{\otimes}(\vec{k}) {\mathcal F}_{\otimes}(\vec{p}) \rangle = \langle {\mathcal F}_{\oplus}(\vec{k}) {\mathcal F}_{\oplus}(\vec{p}) \rangle =
\frac{2 \pi^2}{k^3} {\mathcal P}_{T}(k) \delta^{(3)}(\vec{k} + \vec{p}), 
\label{TPS}
\end{equation}
but $\langle {\mathcal F}_{\otimes}(\vec{k}) {\mathcal F}_{\oplus}(\vec{p}) \rangle =0$.  
The source terms are, in the tensor case:
\begin{eqnarray}
&&C_{I}^{(\mathrm{t})} = \frac{3 \epsilon'}{8 \pi} \int d\Omega'\, \sum_{J = I,\, Q,\, U} {\mathcal N}_{I J} {\mathcal I}^{(\mathrm{t})}_{J} = \epsilon' (1 - f_{e}^2) S_{P}^{(\mathrm{t})}(k,\tau) {\mathcal D}_{T}(\hat{n}) 
\label{CTI}\\
&&C_{\pm}^{(\mathrm{s})} = \frac{3 \epsilon'}{8 \pi} \int d\Omega'\, \sum_{J = I,\, Q,\, U} \biggl[{\mathcal N}_{Q J} \pm i {\mathcal N}_{U J} \biggr]{\mathcal I}^{(\mathrm{t})}_{J} = 
\epsilon' (1 - f_{e}^2) S_{P}^{(\mathrm{t})}(k,\tau){\mathcal D}_{\pm}(\hat{n}),
\label{CTpm}
\end{eqnarray}
where $S_{P}^{(\mathrm{t})}(k,\tau)$ is given by:
\begin{eqnarray}
S_{P}^{(\mathrm{t})}(k,\tau) &=& \frac{3}{32} \int_{-1}^{1} d \mu'
[  (1 - \mu^{\prime 2})^2 \delta_{T}^{(\mathrm{t})}(\mu^{\prime})
- ( 1 + \mu^{\prime 4} + 6 \mu^{\prime 2}) \delta_{P}^{(\mathrm{t})}(\mu^{\prime})] 
\nonumber\\
&=& \frac{3}{70}\delta_{T4}^{(\mathrm{t})} + \frac{\delta_{T2}^{(\mathrm{t})}}{7} - \frac{\delta_{T0}^{(\mathrm{t})}}{10}- \frac{3}{70}\delta_{P4}^{(\mathrm{t})}  
+ \frac{6}{7}\delta_{P2}^{(\mathrm{t})}  - \frac{3}{5}\delta_{P0}^{(\mathrm{t})},
\label{T14}
\end{eqnarray}
and has been computed to lowest order in $X_{F}$, i.e.  $\overline{\delta}_{+}^{(\mathrm{t})}(k,\tau) = \overline{\delta}_{+}^{(\mathrm{t})}(k,\tau) = \delta^{(\mathrm{t})}_{P}(k,\tau)$.
The evolution equations become, therefore,
\begin{eqnarray}
&&  \partial_{\tau} \delta_{T}^{(\mathrm{t})} + ( i k \mu + \epsilon') \delta_{T}^{(\mathrm{t})}= - \frac{h'}{2} + \epsilon' \, ( 1 - f_{e}^2) \,S^{(\mathrm{t})}_{P},
\label{Boltz84}\\
&&  \partial_{\tau} \overline{\delta}^{(\mathrm{t})}_{\pm}+ ( i k \mu + \epsilon') \overline{\delta}^{(\mathrm{t})}_{\pm} = 
- \epsilon' \, ( 1 - f_{e}^2) \, S^{(\mathrm{t})} \mp 2 \,i\, X_{F}(\tau) \, \overline{\delta}^{(\mathrm{t})}_{\pm}.
\label{Boltz85}
\end{eqnarray}
As in the scalar case, (see Eq. (\ref{inhom1})),  Eq. (\ref{Boltz85}) can be generalized to the case where the stochastic process is not spatially 
homogeneous leading, in Fourier space, to an integrodifferential equation containing also a convolution.

 
\renewcommand{\theequation}{5.\arabic{equation}}
\setcounter{equation}{0}
\section{E mode and B mode power spectra}
\label{sec5}
In the present discussion  there are in fact two E mode polarizations
and two B mode polarizations. All the four angular power spectra are affected by the stochastic Faraday rate. 
The two E modes come from  the adiabatic scalar mode and from the tensor fluctuations 
of the geometry: the  scalar E mode induces a B mode 
but also the tensor B mode is modified by the stochastic Faraday effect. The 
total polarization power spectra are related to their scalar and tensor components by means 
of a unitary transformation. To clarify the discussion as much as possible  we shall carefully distinguish
between the scalar and the tensor contributions to the polarization power spectra. 
\subsection{Scalar case}
The brightness perturbations $\Delta^{(\mathrm{s})}_{\pm}(\hat{n},\tau)$ can be expanded in terms of spin-$\pm2$ spherical harmonics $_{\pm 2}Y_{\ell\,m}(\hat{n})$ \cite{EB1,EB2}
\begin{equation}
\Delta^{(\mathrm{s})}_{\pm}(\hat{n},\tau) = \sum_{\ell \, m} a_{\pm 2,\,\ell\, m} \, _{\pm 2}Y_{\ell\, m}(\hat{n}),
\label{int2}
\end{equation}
so that the E and B modes are, up to a sign, the real and the imaginary 
parts of $a_{\pm 2,\ell\,m}$, i.e. 
\begin{equation}
a^{(\mathrm{E})}_{\ell\, m} = - \frac{1}{2}(a_{2,\,\ell m} + a_{-2,\,\ell m}), \qquad  
a^{(\mathrm{B})}_{\ell\, m} =  \frac{i}{2} (a_{2,\,\ell m} - a_{-2,\,\ell m}).
\label{int3}
\end{equation}
In real space the fluctuations constructed from 
$a^{(\mathrm{E})}_{\ell\,m}$ and $a^{(\mathrm{B})}_{\ell\,m}$ have the 
property of being invariant under rotations on a plane orthogonal 
to $\hat{n}$.  They can therefore 
be expanded in terms of spin-0 spherical harmonics:
\begin{equation}
\Delta^{(\mathrm{s})}_{\mathrm{E}}(\hat{n},\tau) = \sum_{\ell\, m} N_{\ell}^{-1} \,  a^{(\mathrm{E})}_{\ell\, m}  \, Y_{\ell\, m}(\hat{n}),\qquad 
\Delta^{(\mathrm{s})}_{\mathrm{B}}(\hat{n},\tau) = \sum_{\ell\, m} N_{\ell}^{-1} \,  a^{(\mathrm{B})}_{\ell\, m}  \, Y_{\ell\, m}(\hat{n}),
\label{int4}
\end{equation}
where $N_{\ell} = \sqrt{(\ell - 2)!/(\ell +2)!}$. In real 
space the scalars $\Delta^{(\mathrm{s})}_{\mathrm{E}}(\hat{n},\tau) $ and $\Delta^{(\mathrm{s})}_{\mathrm{B}}(\hat{n},\tau)$ can be expressed in terms of the generalized 
ladder operators \cite{EB1} raising and lowering the spin-weight of a given function:
\begin{eqnarray}
&& \Delta^{(\mathrm{s})}_{\mathrm{E}}(\hat{n},\tau) = - \frac{1}{2} \{ K_{-}^{(1)}(\hat{n})[K_{-}^{(2)}(\hat{n})
\Delta_{+}(\hat{n},\tau)] +  K_{+}^{(-1)}(\hat{n})[K_{+}^{(-2)}(\hat{n}) \Delta_{-}(\hat{n},\tau)]\},
\label{BE1}\\
&&  \Delta^{(\mathrm{s})}_{\mathrm{B}}(\hat{n},\tau) =  \frac{i}{2} \{ K_{-}^{(1)}(\hat{n})[K_{-}^{(2)}(\hat{n})
\Delta_{+}(\hat{n},\tau)] -  K_{+}^{(-1)}(\hat{n})[K_{+}^{(-2)}(\hat{n}) \Delta_{-}(\hat{n},\tau)]\}.
\label{BE2}
\end{eqnarray}
The differential operators appearing in Eqs. (\ref{BE1}) and (\ref{BE2}) 
are  generalized ladder operators (see \cite{EB1}, first paper) whose action 
either raises or lowers the spin weight of a given fluctuation. They are defined within the present conventions as 
acting on a fluctuation of spin weight $j$: 
\begin{eqnarray}
&& K_{+}^{j}(\hat{n}) = - (\sin{\vartheta})^{j}\biggl[ \partial_{\vartheta} + 
\frac{i}{\sin{\vartheta}} \partial_{\varphi}\biggr] \frac{1}{(\sin{\vartheta})^{j}},
\label{Kp}\\
&& K_{-}^{\mathrm{s}}(\hat{n}) = - \frac{1}{(\sin{\vartheta})^{j}}
\biggl[ \partial_{\vartheta} -
\frac{i}{\sin{\vartheta}} \partial_{\varphi}\biggr] (\sin{\vartheta})^{j}.
\label{Km}
\end{eqnarray}
For instance $K_{-}^{(2)}\Delta_{+}$  transforms as a function of spin-weight 1 while 
$K_{-}^{(1)}[K_{-}^{(2)}\Delta_{+}]$ is, as anticipated, as scalar. Using Eqs. (\ref{Kp}) and (\ref{Km}) 
inside Eqs. (\ref{BE1}) and (\ref{BE2}) the explicit expressions of the E-mode and of the B-mode 
are, in real space:
\begin{eqnarray}
\Delta^{(\mathrm{s})}_{\mathrm{E}} (\hat{n},\tau) &=& - \frac{1}{2}\biggl\{( 1 -\mu^2) \partial_{\mu}^2 (\Delta^{(\mathrm{s})}_{+} + \Delta^{(\mathrm{s})}_{-}) - 4 \mu  \partial_{\mu}(\Delta^{(\mathrm{s})}_{+} + \Delta^{(\mathrm{s})}_{-}) - 2  (\Delta^{(\mathrm{s})}_{+} + \Delta^{(\mathrm{s})}_{-}) 
\nonumber\\
&-& 
\frac{\partial_{\varphi}^2 (\Delta^{(\mathrm{s})}_{+} + \Delta^{(\mathrm{s})}_{-})}{1 - \mu^2 } 
+
2 i\biggl[ \partial_{\varphi}  \partial_{\mu}(\Delta^{(\mathrm{s})}_{+} - \Delta^{(\mathrm{s})}_{-}) - \frac{\mu}{1 - \mu^2} \partial_{\varphi}  (\Delta^{(\mathrm{s})}_{+} - \Delta^{(\mathrm{s})}_{-})\biggr] \biggr\},
\label{BE3}\\
 \Delta^{(\mathrm{s})}_{\mathrm{B}} (\hat{n},\tau) &=& \frac{i}{2} \biggl\{( 1 -\mu^2) \partial_{\mu}^2(\Delta^{(\mathrm{s})}_{+} - \Delta^{(\mathrm{s})}_{-}) - 4 \mu  \partial_{\mu}(\Delta^{(\mathrm{s})}_{+} - \Delta^{(\mathrm{s})}_{-}) - 2  (\Delta^{(\mathrm{s})}_{+} - \Delta^{(\mathrm{s})}_{-})  
\nonumber\\
&-& 
\frac{\partial_{\varphi}^2  (\Delta^{(\mathrm{s})}_{+} - \Delta^{(\mathrm{s})}_{-})}{1 - \mu^2 }
+ 
2 i\biggl[ \partial_{\varphi}  \partial_{\mu}(\Delta^{(\mathrm{s})}_{+} + \Delta^{(\mathrm{s})}_{-}) - \frac{\mu}{1 - \mu^2} \partial_{\varphi}  (\Delta^{(\mathrm{s})}_{+} + \Delta^{(\mathrm{s})}_{-})\biggr] \biggr\}.
\label{BE4}
\end{eqnarray}
The lengthy expressions reported in Eqs. (\ref{BE3}) and (\ref{BE4}) simplify greatly 
since the scalar modes do not have azimuthal dependence:
\begin{eqnarray}
&&\Delta^{(\mathrm{s})}_{E} (\hat{n},\tau) =  - \frac{1}{2} \partial_{\mu}^2\, \biggl[ ( 1 -\mu^2)  \biggl(\Delta^{(\mathrm{s})}_{+} + \Delta^{(\mathrm{s})}_{-}\biggr)\biggr], 
\label{EE}\\
&& \Delta^{(\mathrm{s})}_{B} (\hat{n},\tau) = \frac{i}{2} \partial_{\mu}^2\, \biggl[ ( 1 -\mu^2)  \biggl(\Delta^{(\mathrm{s})}_{+} - \Delta^{(\mathrm{s})}_{-}\biggr)\biggr].
 \label{BB}
 \end{eqnarray}
The angular power spectra for the E mode polarization and for the B mode polarization are then defined as
\begin{equation}
\overline{C}_{\ell}^{(\mathrm{EE})} = \frac{1}{2\ell + 1} \sum_{m = -\ell}^{\ell} 
\langle a^{(\mathrm{E})*}_{\ell m}\,a^{(\mathrm{E})}_{\ell m}\rangle,\qquad 
\overline{C}_{\ell}^{(\mathrm{BB})} = \frac{1}{2\ell + 1} \sum_{m=-\ell}^{\ell} 
\langle a^{(\mathrm{B})*}_{\ell m}\,a^{(\mathrm{B})}_{\ell m}\rangle,
\label{int5}
\end{equation}
where $\langle ...\rangle$ denotes the ensemble average. 
We can finally determine $a_{\ell m}^{(E)}$ and $a_{\ell m}^{(B)}$ within the set of conventions 
followed here:
\begin{eqnarray}
a_{\ell m}^{(E)} &=& - \frac{N_{\ell}}{2 (2 \pi)^{3/2}} \int d \hat{n} \, Y_{\ell m}^{*}(\hat{n})\int  d^{3} k \,  \partial_{\mu}^2 \biggl\{ (1 - \mu^2) \biggl[\delta^{(\mathrm{s})}_{+}(\vec{k},\tau) + \delta_{-}^{(\mathrm{s})}(\vec{k},\tau)\biggr]\biggr\},
\nonumber\\
a_{\ell m}^{(B)} &=&  \frac{i\, \,N_{\ell}}{2 (2 \pi)^{3/2}} \int d \hat{n} \, Y_{\ell m}^{*}(\hat{n})\int  d^{3} k \, \partial_{\mu}^2 \biggl\{ (1 - \mu^2) \biggl[\delta_{+}^{(\mathrm{s})}(\vec{k},\tau) - \delta_{-}^{(\mathrm{s})}(\vec{k},\tau)\biggr]\biggr\}.
\label{cor5}
\end{eqnarray}
The solutions for $\delta_{\pm}^{(\mathrm{s})}$ (see Eq. (\ref{SS})) must then be used inside Eq. (\ref{cor5}). The the obtained expression 
can be averaged over the stochastic process by using Eqs. (\ref{un7a})--(\ref{cor4}).  Thus, if the scalars would be the only source of polarization the angular power spectra of the polarization can then be expressed as
\begin{equation}
C_{\ell}^{(EE)} = e^{- \omega_{F}} \, \cosh{\omega_{F}} \, \overline{C}_{\ell}^{(EE)}, \qquad C_{\ell}^{(BB)} = e^{- \omega_{F}} \, \sinh{\omega_{F}} \, \overline{C}_{\ell}^{(EE)},
\label{int6}
\end{equation}
where $\overline{C}_{\ell}^{(EE)}$ is the E-mode autocorrelation produced
 by the standard adiabatic mode and in the absence of Faraday mixing:
\begin{eqnarray}
\overline{C}_{\ell}^{(EE)} &=& 4\pi \int\frac{d k}{k} \bigl| \Delta_{E\ell}^{(\mathrm{s})}(k,\tau)\bigr|^2,
\label{CCSE1}\\
 \Delta_{E\ell}^{(\mathrm{s})}(k,\tau) &=& \frac{3}{4} \sqrt{\ell (\ell-1) (\ell +1) (\ell +2)} \, \int_{0}^{\tau} {\mathcal K}(\tau_{1})\, S_{P}^{(\mathrm{s})}(k,\tau_{1}) \frac{j_{\ell}(x)}{x^2} d\tau_{1}.
\label{CCSE2}
\end{eqnarray}
In Eq. (\ref{CCSE2}) $j_{\ell}(x)$ are spherical Bessel functions and $x = k (\tau - \tau_{1})$. As already mentioned, the results have been 
derived in the sudden decoupling limit but can be extended to the case where the visibility function has a finite thickness \cite{pav1,pav2,pav3}.
Note, finally, that sometimes it is common to separate $S_{P}^{(\mathrm{s})}(k,\tau_{1}) = \sqrt{{\mathcal P}}_{{\mathcal R}}(k) \overline{S}_{P}^{(\mathrm{s})}(k,\tau_{1})$ where 
${\mathcal P}_{\mathcal R}(k)$ is the power spectrum of the constant adiabatic mode defined as in section \ref{sec2} (see Eq. (\ref{av1a})).
The result of Eq. (\ref{int6}) has been already anticipated in \cite{RC1} and will now be completed by computing the tensor contribution.
  
\subsection{Tensor case}
Also the tensor brightness perturbations $\Delta^{(\mathrm{t})}_{\pm}(\hat{n},\tau)$ can be expanded in terms of spin-$\pm2$ spherical harmonics $_{\pm 2}Y_{\ell\,m}(\hat{n})$, i.e. 
\begin{equation}
\Delta^{(\mathrm{t})}_{\pm}(\hat{n},\tau) = \sum_{\ell \, m} b_{\pm 2,\,\ell\, m} \, _{\pm 2}Y_{\ell\, m}(\hat{n}).
\label{int2a}
\end{equation}
In full analogy with the previous case the E and B modes are defined as
\begin{equation}
b^{(\mathrm{E})}_{\ell\, m} = - \frac{1}{2}(b_{2,\,\ell m} + b_{-2,\,\ell m}), \qquad  
b^{(\mathrm{B})}_{\ell\, m} =  \frac{i}{2} (b_{2,\,\ell m} - b_{-2,\,\ell m}).
\label{int3a}
\end{equation}
The fluctuations constructed from 
$b^{(\mathrm{E})}_{\ell\,m}$ and $b^{(\mathrm{B})}_{\ell\,m}$ have the 
property of being invariant under rotations on a plane orthogonal 
to $\hat{n}$ and they can be be expanded in terms of (ordinary) spherical harmonics:
\begin{equation}
\Delta^{(\mathrm{t})}_{\mathrm{E}}(\hat{n},\tau) = \sum_{\ell\, m} N_{\ell}^{-1} \,  b^{(\mathrm{E})}_{\ell\, m}  \, Y_{\ell\, m}(\hat{n}),\qquad 
\Delta^{(\mathrm{t})}_{\mathrm{B}}(\hat{n},\tau) = \sum_{\ell\, m} N_{\ell}^{-1} \,  b^{(\mathrm{B})}_{\ell\, m}  \, Y_{\ell\, m}(\hat{n}),
\label{int4a}
\end{equation}
where $N_{\ell} = \sqrt{(\ell - 2)!/(\ell +2)!}$.  From Eqs. (\ref{BE1}) and (\ref{BE2}) written in the tensor case we can also deduce $b^{(\mathrm{E})}_{\ell m}$ and $b^{(\mathrm{B})}_{\ell m}$ and, in particular, we shall have 
\begin{eqnarray}
b_{\ell m}^{(E)} &=& \frac{N_{\ell}}{(2 \pi)^{3/2}} \int d \hat{n} \, Y_{\ell m}^{*}(\hat{n})\int  d^{3} k \,  \Delta_{E}^{(\mathrm{t})}(k,\hat{n},\tau),
\nonumber\\
b_{\ell m}^{(B)} &=&  \frac{N_{\ell}}{(2 \pi)^{3/2}} \int d \hat{n} \, Y_{\ell m}^{*}(\hat{n})\int  d^{3} k \,  \Delta_{B}^{(\mathrm{t})}(k,\hat{n},\tau).
\label{cort5}
\end{eqnarray}
After some algebra $ \Delta_{E}^{(\mathrm{t})}(\tau,\vec{k},\hat{n})$ and $\Delta_{B}^{(\mathrm{t})}(\tau,\vec{k},\hat{n})$ can be expressed as:
\begin{eqnarray}
&\Delta^{(\mathrm{t})}_{\mathrm{E}}(k,\hat{n},\tau) = q_{+}(\hat{n}) \Lambda_{E}(x)\, {\mathcal Y}_{+}(k,x,\tau,\tau_{1})  - q_{-}(\hat{n}) \Lambda_{B}(x){\mathcal Y}_{-}(k,x,\tau,\tau_{1}) ,
\label{cort6}\\
&\Delta^{(\mathrm{t})}_{\mathrm{B}}(k,\hat{n},\tau) = q_{+}(\hat{n}) \Lambda_{E}(x) \,{\mathcal Y}_{-}(k,x,\tau,\tau_{1}) - q_{-}(\hat{n}) \Lambda_{B}(x){\mathcal Y}_{+}(k,x,\tau,\tau_{1}),
\label{cort7}
\end{eqnarray}
where $x = k (\tau - \tau_{1})$; the functions ${\mathcal Y}_{\pm}(k,x,\tau,\tau_{1})$ and $q_{\pm}(\hat{n})$ are defined as
\begin{eqnarray}
{\mathcal Y}_{\pm}(k,x,\tau,\tau_{1}) &=&\frac{1}{2} \int_{0}^{\tau} e^{- i \mu x} {\mathcal K}^{(\mathrm{t})}(\tau_{1}) \, S_{P}^{(\mathrm{t})}(k,\tau_{1})\biggl[ {\mathcal A}_{+}(\tau,\tau_{1}) 
\pm {\mathcal A}_{-}(\tau,\tau_{1})\biggr]\, d\tau_{1},
\label{YY}\\
q_{\pm}(\hat{n}) &=& (1-\mu^2) \biggl[ e^{ 2 i \varphi} {\mathcal F}_{1}(\vec{k}) \pm e^{- 2 i \varphi} {\mathcal F}_{2}(\vec{k}) \biggr].
\label{qq}
\end{eqnarray}
Furthermore $\Lambda_{E}(x) $ and $\Lambda_{B}(x)$ are two differential operators 
\begin{equation}
\Lambda_{E}(x) = -12+ x^2( 1 - \partial_{x}^2) - 8 x \partial_{x}, \qquad \Lambda_{B}(x) = 8 x + 2 x^2 \partial_{x}.
\end{equation}
The EE and BB angular power spectra are then defined as 
\begin{equation}
C_{\ell}^{(\mathrm{EE})} = \frac{1}{2\ell + 1} \sum_{m = -\ell}^{\ell} 
\langle b^{(\mathrm{E})*}_{\ell m}\,b^{(\mathrm{E})}_{\ell m}\rangle,\qquad 
 C_{\ell}^{(\mathrm{BB})} = \frac{1}{2\ell + 1} \sum_{m=-\ell}^{\ell} 
\langle b^{(\mathrm{B})*}_{\ell m}\,b^{(\mathrm{B})}_{\ell m}\rangle,
\label{int5a}
\end{equation}
As in the scalar case, using Eqs. (\ref{un7a})--(\ref{cor4}) and  the results of section Appendix \ref{APPC}, the stochastic averages can be computed and 
 the angular power spectra of the polarization can then be expressed as
\begin{eqnarray}
&& C_{\ell}^{(EE)} = e^{- \omega_{F}} \, \cosh{\omega_{F}} \, {\mathcal C}_{\ell}^{(EE)} + e^{-\omega_{F}} \,  \sinh{\omega_{F}} \, {\mathcal C}_{\ell}^{(BB)} , 
\nonumber\\
&& C_{\ell}^{(BB)} = e^{- \omega_{F}} \, \sinh{\omega_{F}} \, {\mathcal C}_{\ell}^{(EE)} +  e^{-\omega_{F}} \,  \cosh{\omega_{F}} \, {\mathcal C}_{\ell}^{(BB)},
\label{int6a}
\end{eqnarray}
where ${\mathcal C}_{\ell}^{(EE)}$ and ${\mathcal C}_{\ell}^{(B)}$ (in calligraphic style) are the E-mode autocorrelation produced
 by the standard adiabatic mode and in the absence of Faraday mixing:
 \begin{eqnarray}
 {\mathcal C}_{\ell}^{(EE)} &=& 4 \pi \int \frac{d k}{k} \bigl| \Delta_{E\ell}^{(\mathrm{t})}(k,\tau)\bigr|^2,\qquad {\mathcal C}_{\ell}^{(BB)} = 4 \pi \int \frac{d k}{k} \bigl| \Delta_{B\ell}^{(\mathrm{t})}(k,\tau)\bigr|^2,
 \label{CCT1}\\
 \Delta_{E\ell}^{(\mathrm{t})}(k,\tau) &=& \int_{0}^{\tau} d\tau_{1} {\mathcal K}(\tau_{1}) S_{P}^{(\mathrm{t})}(k,\tau_{1}) \, \Lambda_{E}(x) \frac{j_{\ell}(x)}{x^2}, 
\label{CCT2}\\
 \Delta_{B\ell}^{(\mathrm{t})}(k,\tau) &=& \int_{0}^{\tau} d\tau_{1} {\mathcal K}(\tau_{1}) S_{P}^{(\mathrm{t})}(k,\tau_{1}) \, \Lambda_{B}(x) \frac{j_{\ell}(x)}{x^2}.
\label{CCT3}
\end{eqnarray}
As in the scalar case the results have been derived in the sudden decoupling limit \cite{pav1,pav2,pav3}. Furthermore, as in the scalar case it is possible 
to separate the polarization source as $ S_{P}^{(\mathrm{t})}(k,\tau_{1}) = \sqrt{{\mathcal P}_{T}(k)} \overline{S}_{P}^{(\mathrm{t})}(k,\tau_{1})$ where 
${\mathcal P}_{T}(k)$ is the tensor power spectrum defined as in Eq. (\ref{TPS}) and related to the scalar power spectrum as ${\mathcal P}_{T}(k) = r_{T} {\mathcal P}_{\mathcal R}$ 
where $r_{T}$ is the tensor to scalar ratio measuring the ratio between the scalar and tensor contributions at the pivot scale $k_{p}$ (see also Eq. (\ref{av1a})).
Equations (\ref{int6a}) hold under the academic hypothesis that the tensors are be the only source of polarization. In the realistic case Eqs. (\ref{int6}) and (\ref{int6a}) 
shall be modified even further. 
\subsection{Total polarization power spectra} 
If the scalar and tensor modes of the geometry are simultaneously present the total angular power spectra are obtained 
from Eqs. (\ref{int6}) and (\ref{int6a}) and the result is:
\begin{eqnarray}
C_{\ell}^{(EE)} &=& e^{-\omega_{F}} \cosh{\omega_{F}} \, \biggl( \overline{C}_{\ell}^{(EE)} + {\mathcal C}_{\ell}^{(EE)} \biggr) + e^{- \omega_{F}}\, \sinh{\omega_{F}}\,
 {\mathcal C}_{\ell}^{(BB)},
 \nonumber\\
 {\mathcal C}_{\ell}^{(BB)} &=& e^{-\omega_{F}} \sinh{\omega_{F}} \biggl( \overline{C}_{\ell}^{(EE)} + {\mathcal C}_{\ell}^{(EE)} \biggr) + e^{-\omega_{F}} \, \cosh{\omega_{F}}\,{\mathcal C}_{\ell}^{(BB)}.
 \label{int7a}
 \end{eqnarray}
As already mentioned in the introduction $C_{\ell}^{(EE)}$ and $C_{\ell}^{(BB)}$ denote, respectively, the total E mode and B mode power spectra,
while $\overline{C}_{\ell}^{(EE)}$ is the E mode coming from the adiabatic scalar mode.
The calligraphic power spectra ${\mathcal C}_{\ell}^{(EE)}$ and  ${\mathcal C}_{\ell}^{(BB)}$ denote, respectively, 
the angular power spectra coming from the tensor modes. The explicit expressions of $\omega_{F}$ can be rather different but the frequency dependence will always be the same: since $\omega_{F}$ is quadratic in the rates it will always scale as $1/\overline{\nu}^{4} \simeq \lambda^{4}$ where $\lambda$ denotes the wavelength of the channel. Since the scale factor is normalized in such a way that $a_{0}=1$, physical and comoving 
frequencies coincide today but not in the past.  Equation (\ref{int7a}) does not assume that the rotation rate is perturbative.  
In the limit $\omega_{F} \ll 1$ Eq. (\ref{int7a}) becomes:
\begin{equation}
C_{\ell}^{(EE)} \simeq \, {\mathcal C}_{\ell}^{(EE)} +\overline{C}_{\ell}^{(EE)} + \omega_{F} {\mathcal C}_{\ell}^{(BB)}, \qquad  {\mathcal C}_{\ell}^{(BB)} \simeq \omega_{F}  
\biggl(\overline{C}_{\ell}^{(EE)} + {\mathcal C}_{\ell}^{(EE)} \biggr)  + {\mathcal C}_{\ell}^{(BB)}.
\label{int7b}
\end{equation}
Equations (\ref{int7a}) and (\ref{int7b}) imply that the E mode coming from the scalar adiabatic 
fluctuations can be turned into a B mode. However, if a tensor B mode is present,  the total angular power spectrum of the E mode is also 
affected. Finally, in the limit $\omega_{F}\to 0$ there is no stochastic mixing and the standard situation is recovered: both the scalar and tensor 
contributions to the E mode and only the tensor contribution to the total B mode power spectrum.

\subsection{Scaling relations and sum rules}
In Ref. \cite{RC1} some 
relations have been derived in the absence of the tensor B mode polarization and  it was  argued that these relations 
could be modified if the tensor B mode was included from the very beginning. We are now in a position to show that 
this conclusion is incorrect since the relations obtained in \cite{RC1} are preserved even in the presence of the tensor B mode.
The reason for this simple result can be understood, a posteriori, from the unitary transformation connecting the total angular power spectra to their 
scalar and tensor components.

Let us therefore show that the nonlinear combinations proposed in \cite{RC1}, in the absence of the tensor B mode, hold also 
in the present case. Defining the properly normalized 
total power spectra of the E mode and B mode polarizations:
\begin{equation}
{\mathcal G}_{E\ell} = \frac{\ell(\ell+1)}{2\pi} C_{\ell}^{(EE)}, \qquad {\mathcal G}_{B\ell} = \frac{\ell(\ell+1)}{2\pi} C_{\ell}^{(BB)}. 
\label{int8a}
\end{equation}
the first nonlinear combination guessed in Ref. \cite{RC1} was:
\begin{equation}
{\mathcal L}_{0}(\omega_{F}) = \frac{{\mathcal G}^2_{E\ell}(\omega_{F}) - {\mathcal G}^2_{B\ell}(\omega_{F})}{[{\mathcal G}_{E\ell}(\omega_{F}) + {\mathcal G}_{B\ell}(\omega_{F})]^2}\to e^{- 2 \omega_{F}}.
\label{int9a}
\end{equation}
The result expressed by Eq. (\ref{int9a}) holds, indeed, also when ${\mathcal G}_{E\ell}$ and ${\mathcal G}_{B\ell}$ are constructed in terms 
of the spectra of Eq. (\ref{int7a}). To demonstrate this point it suffices to insert  Eq. (\ref{int7a})
into Eq. (\ref{int8a}) and into the definition of ${\mathcal L}_{0}(\omega_{F})$ given in the first equality of 
Eq. (\ref{int9a}).  One can easily think of two further combinations with well defined scaling properties with $\omega_{F}$, namely:
\begin{eqnarray} 
{\mathcal L}_{1}(\omega_{F}) &=& \frac{{\mathcal G}_{E\ell}(\omega_{F})  - {\mathcal G}_{B\ell}(\omega_{F})}{{\mathcal G}_{E\ell}(\omega_{F}) +{\mathcal G}_{B\ell}(\omega_{F}) } \to e^{ - 2 \omega_{F}},
\label{int10a}\\
{\mathcal L}_{2}(\omega_{F}) &=& \frac{{\mathcal G}^2_{E\ell}(\omega_{F})  + {\mathcal G}^2_{B\ell}(\omega_{F})}{{\mathcal G}^2_{E\ell}(\omega_{F}) - {\mathcal G}_{B\ell}^2(\omega_{F}) } \to  \cosh{2\omega_{F}}.
\label{int11a}
\end{eqnarray}
As in the case of Eq. (\ref{int9a}), also Eqs. (\ref{int10a}) and (\ref{int11a}) can be directly verified by direct use of Eqs. (\ref{int7a}) and (\ref{int8a}).
Different nonlinear combinations can be invented either by combining nonlinearly ${\mathcal L}_{0}(\omega_{F})$, ${\mathcal L}_{1}(\omega_{F})$ and 
${\mathcal L}_{2}(\omega_{F})$ or by concocting new variables. 
\subsection{Mixing of the B modes}
 It would be nice to have definite criteria in order to rule out or to rule in the explanation of the detected 
B mode in terms of Faraday rotation. In this respect there are two possible criteria. One is to 
look for the frequency scaling and the other is to look for the scaling with the multipoles. Both properties 
are well understood analytically and numerically. In this perspective we find that the existing bounds on the 
B mode polarization at low \cite{dasi,cbi}, intermediate \cite{capmap} and high frequencies \cite{boom,quad} should probably be revisited. Here 
we just list some simple considerations.

The idea, in short, goes as follows. Let us suppose that one or more experiments measures both the E mode and the B mode polarizations 
in different frequency channels. If there is mixing between the B mode polarization of tensor 
origin and the B mode induced directly by Faraday rotation, then, according to Eq. (\ref{int7a})  various combinations 
with definite scaling properties with the comoving frequency can be constructed and some examples have been 
listed in  Eqs. (\ref{int9a}), (\ref{int10a}) and (\ref{int11a}).
If the E mode and the B mode autocorrelations are independently measured in each frequency channel of a given experiment, both scale-invariant and scale-dependent combinations of the angular power spectra can be constructed frequency by frequency. 
So, for instance the combination 
\begin{equation}
\lim_{\omega_{F}\gg 1} = {\mathcal L}_{0} + {\mathcal L}_{2} \to 2,
\label{int12a}
\end{equation}
is asymptotically frequency-independent while the combination 
\begin{equation}
{\mathcal L}_{2}/({\mathcal L}_{0} + {\mathcal L}_{0}^{-1})
\label{int13a}
\end{equation}
is truly frequency-independent. Equations (\ref{int9a}), (\ref{int10a}) and (\ref{int11a}) (or any other nonlinear combination of the power spectra 
constructed with the same criteria) represent a  set of physical observables that can be used to discriminate between the frequency dependence induced by the stochastic Faraday mixing or by other concurrent forms of frequency scaling caused either by the known or by the yet unknown foregrounds.  

In this respect we ought to conclude this section with an extremely relevant but simple comment. If 
the BB correlation comes entirely from the tensor modes of the geometry, the internal linear combination (ilc) technique can be applied indifferently for all the channels 
of the experiment that eventually detect the B mode. In practice, the ilc map is  
a weighted linear combination over the smoothed maps obtained from each of the different frequency channels.
Conversely, if the signal contains a frequency-dependent part, as the Faraday mixing would predict, the ilc technique cannot be applied. 
In these circumstances, the scaling relations and the sum rules obtained here  
are crucial if we intend to disentangle the real physical effects from potential foregrounds.

\renewcommand{\theequation}{6.\arabic{equation}}
\setcounter{equation}{0}
\section{Concluding remarks}
\label{sec6}
This paper investigated the Faraday effect of the CMB as a different and more mundane source of the B mode 
polarization detected by Bicep2. In the first part of the paper we discussed a maximalist alternative to the tensor 
B mode where the whole Bicep2 data are explained by a Faraday rotated E mode polarization. In the second 
part of the paper we discussed the possibility where the tensor B mode interferes with the Faraday rotated 
E mode polarization.

It has been shown both analytically and numerically that the Faraday rotation alone cannot explain the Bicep2 data. If this 
happens other CMB observables will be excessively distorted. The first estimate can be obtained by maximizing the 
E mode autocorrelation and by computing the induced B mode polarization. In this case we see that, given the 
Bicep2 frequency (i.e. $150$ GHz), the Bicep2 normalization can only be reproduced when the 
magnetic field is ${\mathcal O}(15)$ nG. This value of the magnetic field is too large since it would induce unobserved 
distortions in the temperature autocorrelations. Indeed much lower magnetic fields (i.e. ${\mathcal O}(1.5)$ nG ) 
already produce excessive distortions on the TT correlations if the magnetic power spectrum is nearly scale-invariant 
(i.e. $n_{B} \to 1$).  An independent test, in this respect, 
is provided by the frequency scaling of the signal which can be separately discussed. Other signals 
of B mode polarization can be induced directly by the tensor and vector modes of the geometry 
induced by the magnetic fields. It is however well established that these signals are much smaller 
than the Faraday effect for two reasons. First the magnetized adiabatic mode of scalar origin
dominates at the level of the initial conditions and at the level of the TT correlation: the vector 
and tensor modes of magnetic origin are comparatively smaller. Second the B mode signals 
induced by the vectors and the tensors are quadratic in the magnetic power spectrum while 
the Faraday B mode is linear in the magnetic power spectrum.

The realistic situation is therefore the one where there are two 
physically plausible sources of B mode polarization: the first is given by the tensor modes of the geometry, i.e.
relic gravitons with present wavelengths comparable with the Hubble radius; the second is given the Faraday 
rotated E mode polarization. The second part of the present paper dealt with the possibility that 
two effects can interfere.  The B mode polarization of tensor origin is virtually frequency independent. Conversely 
the Faraday rotated E mode polarization does depend on the frequency. 

 Elaborating on a recent suggestion the Faraday effect has been treated as a random, stationary and quasi-Markovian process. 
The stochastic treatment of this phenomenon bears some analogy with the case of synchrotron emission and the obtained 
results encompass and complement previous analyses where the formation of the Faraday effect has been customarily presented 
as a purely deterministic process in time.  Within this approach a set of scaling laws only involving observable 
power spectra can be derived. These scaling laws, once applied to observational data at different 
frequencies, can be used to disentangle the Faraday induced B mode polarization from the tensor 
B mode.

\section*{Acknowledgements}

It is a pleasure to thank A. Gentil-Beccot and S. A. Rohr of the CERN scientific information service for their kind assistance. 

\newpage

\begin{appendix}
\renewcommand{\theequation}{A.\arabic{equation}}
\setcounter{equation}{0}
\section{Description of the polarization}
\label{APPA}
To avoid lengthy digressions, in this Appendix we are going to report some technical details that  can be useful for the 
interested reader. In general terms the brightness perturbations can be arranged in 
a column matrix, be it ${\mathcal I}=(\Delta_{I}, \, \Delta_{Q},\, \Delta_{U}) $ obeying the following equation
\begin{equation}
\frac{d {\mathcal I}}{d\tau} + \epsilon' {\mathcal I} - 2 X_{F}(\vec{x},\tau) {\mathcal W}\,{\mathcal I} = \frac{3 \epsilon'}{ 8 \pi} \int \, {\mathcal N}(\Omega, \Omega') \, {\mathcal I}(\Omega')\, d\Omega' ,
\label{AP1}
\end{equation}
where $\Omega$ denotes the angular dependence and $d \Omega' = d\varphi'\, d\mu'$ with $\mu'= \cos{\vartheta'}$; all the matrix elements of ${\mathcal W}$ vanish except two: ${\mathcal W}_{Q\, U} = -1$ and ${\mathcal W}_{U\, Q} = 1$.
The first term generically denotes the collisionless contribution while the collision term contains the matrix ${\mathcal N}(\Omega,\Omega')$. The matrix elements entering the collision term of the equation for $\Delta_{I}$ are:
\begin{eqnarray}
{\mathcal N}_{I\,I} &=& \frac{1}{4} \biggl[ 3 - \mu^2 - \mu^{\prime 2} + 3 \mu^2 \mu^{\prime 2} + 4 \mu \mu^{\prime} \sqrt{1 - \mu^2} 
\sqrt{1 - \mu^{\prime 2}} \cos{ (\varphi' - \varphi)} 
\nonumber\\
&+& (1 - \mu^2) ( 1 - \mu^{\prime 2}) \cos{2 (\varphi' - \varphi)} \biggr],
\nonumber\\
{\mathcal N}_{I\,Q} &=& \frac{1}{4} \biggl[ (3 \mu^2 -1)(\mu^{\prime 2} -1) + 4 \mu \sqrt{1 - \mu^2} \sqrt{1 - \mu^{\prime 2}} \cos{(\varphi' - 
\varphi)}
\nonumber\\
&+& (\mu^{\prime 2} + 1 ) (\mu^2 -1) \cos{2 (\varphi' - \varphi)}\biggr],
\nonumber\\
{\mathcal N}_{I\,U} &=& - \biggl[ \mu \sqrt{1 - \mu^2} \sqrt{1 - \mu^{\prime 2}} + ( \mu^2 -1) \mu^{\prime} \cos{(\varphi' - \varphi)} \biggr] \sin{(\varphi' - \varphi)};
\label{AP2}
\end{eqnarray} 
the matrix elements entering the collision integral of $\Delta_{Q}$ and $\Delta_{U}$ are:
\begin{eqnarray}
{\mathcal N}_{Q\,I} &=& \frac{1}{4} \biggl[ (\mu^2 -1) (3 \mu^{\prime 2} -1) + 4 \mu \mu^{\prime} \sqrt{1 - \mu^2} \sqrt{1 - \mu^{\prime 2}} 
\cos{(\varphi' - \varphi)}
\nonumber\\
&+& (\mu^2 +1 ) (\mu^{\prime 2} -1) \cos{2(\varphi' - \varphi)} \biggr],
\nonumber\\
{\mathcal N}_{Q\,Q} &=& \frac{1}{4} \biggr[3 (\mu^2 -1) (\mu^{\prime 2} -1) + 4 \mu \mu^{\prime} \sqrt{1 - \mu^{\prime 2}} \sqrt{1 - \mu^2} \cos{(\varphi' - \varphi)} 
\nonumber\\
&+& ( 1 + \mu^2) (1 + \mu^{\prime 2}) \cos{2 (\varphi' - \varphi)}\biggr],
\nonumber\\
{\mathcal N}_{Q\,U} &=&  - \biggl[ \mu \sqrt{1 - \mu^2} \sqrt{1 - \mu^{\prime 2}} + ( \mu^2 +1) \mu^{\prime} \cos{(\varphi' - \varphi)} \biggr] \sin{(\varphi' - \varphi)},
\nonumber\\
{\mathcal N}_{U\, I} &=&\biggl[ \mu^{\prime} \sqrt{1 - \mu^2} \sqrt{1 - \mu^{\prime 2}} + ( \mu^{\prime 2} -1) \mu \cos{(\varphi' - \varphi)} \biggr] \sin{(\varphi' - \varphi)},
\nonumber\\
{\mathcal N}_{U\, Q} &=& \biggl[ \mu^{\prime} \sqrt{1 - \mu^2} \sqrt{1 - \mu^{\prime 2}} + ( \mu^{\prime 2} +1) \mu \cos{(\varphi' - \varphi)} \biggr] \sin{(\varphi' - \varphi)},
\nonumber\\
{\mathcal N}_{U\, U} &=& \sqrt{1 -\mu^2} \sqrt{1 - \mu^{\prime 2}} \cos{(\varphi' - \varphi)} + \mu \mu^{\prime} \cos{2 (\varphi' - \varphi)}.
\label{AP3}
\end{eqnarray}

\renewcommand{\theequation}{B.\arabic{equation}}
\setcounter{equation}{0}
\section{Scalar and tensor fluctuations}
\label{APPB}
The scalar, vector and tensor components of the brightness perturbations are affected, respectively, by the scalar, vector and tensor inhomogeneties of the geometry and of the various sources. It must be however stressed that the fluctuations of the geometry affect directly only the evolution of $\Delta_{I}$ while the linear polarizations 
are indirectly affected by the fluctuations of the geometry through the source terms of the corresponding equations. With these caveats in mind we are first 
going to deduce the collisionless part of the evolution of the brightness perturbations.

Assuming the conformally flat background introduced in 
Eq. (\ref{metric}), the fluctuations of the metric can be written, in general terms, as 
\begin{equation}
\delta g_{\mu\nu}(\vec{x},\tau) = \delta_{\mathrm{s}} g_{\mu\nu}(\vec{x},\tau) 
+ \delta_{\mathrm{v}} g_{\mu\nu}(\vec{x},\tau) + 
 \delta_{\mathrm{t}} g_{\mu\nu}(\vec{x},\tau).
\label{BRDEC1a}
\end{equation}
where $\delta_{\mathrm{s}}$, $\delta_{\mathrm{v}}$ and $\delta_{\mathrm{t}}$ denote 
the inhomogeneity preserving, separately, the scalar, vector and tensor nature of the fluctuations. The scalar modes of the geometry are parametrized in terms of four independent functions. In the longitudinal gauge the scalar fluctuations of the metric are 
\begin{eqnarray}
&& \delta_{\mathrm{s}} g_{00}(\vec{x},\tau) = 2 a^2(\tau) \phi(\vec{x},\tau),\qquad  \delta_{\mathrm{s}} g_{ij}(\vec{x},\tau) = 2 a^2(\tau) \psi(\vec{x},\tau) \delta_{ij}.
\label{BRDEC2}
\end{eqnarray}
The vector modes are described by two independent vectors $Q_{i}(\vec{x},\tau)$ and $W_{i}(\vec{x},\tau)$ 
\begin{equation}
 \delta_{\mathrm{v}} g_{0i}(\vec{x},\tau) = - a^2 Q_{i}(\vec{x},\tau),\qquad \delta_{\mathrm{v}} g_{ij}(\vec{x},\tau) = a^2 \biggl[\partial_{i} W_{j}(\vec{x},\tau) + \partial_{j}W_{i}(\vec{x},\tau)\biggr],
\label{BRDEC3}
\end{equation}
subjected to the conditions $\partial_{i} Q^{i} =0$ and $\partial_{i} W^{i} =0$. Also in the vector case it is possible to fix a gauge by choosing, for instance, $Q_{i} =0$.
Finally the tensor  modes of the geometry are 
parametrized in terms of a rank-two tensor in three spatial dimensions, i.e.
\begin{equation}
\delta_{t} g_{ij}(\vec{x},\tau) = - a^2 h_{ij}, \qquad \partial_{i} h^{i}_{j}(\vec{x},\tau) = h_{i}^{i}(\vec{x},\tau) = 0,
\label{BRDEC4}
\end{equation}
which is automatically invariant under infinitesimal coordinate transformations.  

Ignoring, for the moment, the collision terms we have that the collisionless 
parts of the evolution of $\Delta_{I}$ can be written as
\begin{eqnarray}
&& {\mathcal L}_{I}^{(\mathrm{s})}(\hat{n},\vec{x},\tau) = \partial_{\tau} \Delta^{(\mathrm{s})}_{I}  + \hat{n}^{i} \partial_{i} \Delta^{(\mathrm{s})}_{I} + \epsilon' \Delta^{(\mathrm{s})}_{I} + \frac{1}{q} \biggl(\frac{d q}{d\tau}\biggr)_{\mathrm{s}},
\label{BRDEC6}\\
&& {\mathcal L}_{I}^{(\mathrm{v})}(\hat{n}, \vec{x},\tau) = \partial_{\tau} \Delta^{(\mathrm{v})}_{I} + \hat{n}^{i} \partial_{i} \Delta^{(\mathrm{v})}_{I} + \epsilon' \Delta^{(\mathrm{v})}_{I} + \frac{1}{q}\biggl(\frac{d q}{d\tau}\biggr)_{\mathrm{v}},
\label{BRDEC7}\\
&&  {\mathcal L}_{I}^{(\mathrm{t})}(\hat{n},\vec{x},\tau) = \partial_{\tau} \Delta^{(\mathrm{t})}_{I} + \hat{n}^{i} \partial_{i} \Delta^{(\mathrm{t})}_{I} + \epsilon' \Delta^{(\mathrm{t})}_{I} + \frac{1}{q} \biggl(\frac{d q}{d\tau}\biggr)_{\mathrm{t}},
\label{BRDEC8}
\end{eqnarray}
where $q = \hat{n}_{i} q^{i}$ denotes the comoving three-momentum. The scalar, vector and tensor contributions to the derivatives of the modulus of the comoving three-momentum are given, respectively, by 
\begin{eqnarray}
&& \biggl(\frac{d q}{d\tau}\biggr)_{\mathrm{s}} = - q \partial_{\tau} \psi + q \hat{n}^{i} \partial_{i} \phi, 
\label{BRDEC9a}\\
&& \biggl(\frac{d q}{d\tau}\biggr)_{\mathrm{v}} = \frac{q}{2} \hat{n}^{i} \hat{n}^{j} (\partial_{i} \partial_{\tau}W_{j} + \partial_{\tau}\partial_{j} W_{i}),
\label{BRDEC9}\\
&& \biggl(\frac{d q}{d\tau}\biggr)_{\mathrm{t}} = - \frac{q}{2} \, \hat{n}^{i}\, \hat{n}^{j} \, \partial_{\tau}h_{ij}.
\label{BRDEC10}
\end{eqnarray}
 The notation introduced in Eqs. (\ref{BRDEC6}), 
 (\ref{BRDEC7}), (\ref{BRDEC8}) for the fluctuations of the intensity can also 
 be generalized to the linear and circular polarizations: 
 \begin{equation}
 {\mathcal L}^{(\mathrm{y})}_{X}(\hat{n},\vec{x},\tau) = \partial_{\tau} \Delta^{(\mathrm{y})}_{X} + \hat{n}^{i} \partial_{i} \Delta^{(\mathrm{y})}_{X}  + \epsilon' \Delta^{(\mathrm{y})}_{X},
\label{BRDEC11a}
\end{equation}
where the subscript can coincide, alternatively, with $Q$, $U$ and $V$ (i.e. $X= Q,\, U,\, V$) and the superscript denotes the 
transformation properties of the given fluctuation (i.e. $\mathrm{y} = \mathrm{s},\,\mathrm{v},\,\mathrm{t}$). 

The vector and the tensor polarizations can be decomposed, respectively, as
\begin{eqnarray}
&& W_{i}(\vec{k},\tau) = \sum_{\lambda} e^{(\lambda)}_{i} W_{(\lambda)}(\vec{k},\tau) = \hat{a}_{i} W_{a}(\vec{k},\tau) + 
\hat{b}_{i} W_{b}(\vec{k},\tau),
\label{BRDEC14a}\\
&& h_{ij}(\vec{k},\tau) =\sum_{\lambda} \epsilon^{(\lambda)}_{ij} h_{(\lambda)}(\vec{k},\tau)  = \epsilon^{\oplus}_{ij} 
h_{\oplus}(\vec{k},\tau) + \epsilon^{\otimes}_{ij} h_{\otimes}(\vec{k},\tau),
\label{BRDEC14b}
\end{eqnarray}
where $\hat{k}$ denotes the direction of propagation and
 the two orthogonal directions $\hat{a}$ and $\hat{b}$ are such that $\hat{a}\times \hat{b} = \hat{k}$. Given the direction of propagation of the relic tensor 
 oriented along $\hat{k}$,  
the two tensor polarizations are defined in terms of $\hat{a}_{i}$ and $\hat{b}_{i}$ as:
\begin{equation}
\epsilon^{\oplus}_{ij}(\hat{k}) = \hat{a}_{i} \hat{a}_{j} - \hat{b}_{i} \hat{b}_{j}, \qquad 
\epsilon^{\otimes}_{ij}(\hat{k}) = \hat{a}_{i} \hat{b}_{j} + \hat{a}_{j} \hat{b}_{i}.
\label{BRDEC17}
\end{equation}
The projections of the vector and of the tensor polarizations on the direction of photon propagation $\hat{n}$ are:
\begin{eqnarray}
&&\hat{n}^{i} W_{i}(\vec{k},\tau)=  \biggl[ \hat{n}^{i} \hat{a}_{i} W_{a}(\vec{k},\tau) + \hat{n}^{i}\hat{b}_{i} W_{b}(\vec{k},\tau)\biggr],
\label{BRDEC15}\\
&&\hat{n}^{i} \hat{n}^{j} h_{i j}(\vec{k},\tau) =  \biggl\{ [ (\hat{n}\cdot\hat{a})^2 
- (\hat{n}\cdot\hat{b})^2] h_{\oplus}(\vec{k},\tau) + 2 (\hat{n}\cdot\hat{a}) (\hat{n} \cdot \hat{b})
h_{\otimes}(\vec{k},\tau)\biggr\}.
\label{BRDEC16}
\end{eqnarray}
Choosing the direction of propagation of the relic vector and of the relic tensor 
along the $\hat{z}$ axis, the unit vectors 
$\hat{a}$ and $\hat{b}$ will coincide with the remaining two Cartesian directions
and the related Fourier amplitudes will satisfy 
\begin{eqnarray}
&& \hat{n}^{i} W_{i}(\vec{k},\tau) = \sqrt{\frac{2\pi}{3}} \biggl[ W_{L}(\vec{k},\tau) \, Y_{1}^{-1}(\vartheta, \varphi) - W_{R}(\vec{k},\tau) Y_{1}^{1}(\vartheta, \varphi)\biggr] ,
\label{BRDEC18}\\
&& \hat{n}^{i} \hat{n}^{j} h_{i j}(\vec{k},\tau) =  \biggl[ h_{R}(\vec{k},\tau) 
Y_{2}^{2}(\vartheta, \varphi) +  h_{L}(\vec{k},\tau) Y_{2}^{-2}(\vartheta, \varphi)\biggr],
\label{BRDEC19}
\end{eqnarray}
where 
\begin{eqnarray}
&& W_{L}(\vec{k},\tau) = \frac{W_{a}(\vec{k},\tau) + i W_{b}(\vec{k},\tau)}{\sqrt{2}},\qquad W_{R}(\vec{k},\tau) = \frac{W_{a}(\vec{k},\tau) - i W_{b}(\vec{k},\tau)}{\sqrt{2}},
\nonumber\\
&& h_{L}(\vec{k},\tau) = \frac{h_{\oplus}(\vec{k},\tau) + i h_{\otimes}(\vec{k},\tau)}{\sqrt{2}},\qquad h_{R}(\vec{k},\tau) = \frac{h_{\oplus}(\vec{k},\tau) - i h_{\otimes}(\vec{k},\tau)}{\sqrt{2}};
\label{BRDEC20}
\end{eqnarray}
the spherical harmonics appearing in Eqs. (\ref{BRDEC18}) and (\ref{BRDEC19}) 
are, respectively, 
\begin{equation}
Y_{1}^{\pm 1} (\vartheta,\varphi) = \mp \sqrt{\frac{3}{8\pi}} \, \sin{\vartheta} \,\,e^{\pm i \varphi},\qquad Y_{2}^{\pm 2} (\vartheta,\varphi) = \sqrt{\frac{15}{32 \pi}} \sin^2{\vartheta} \,\,e^{ \pm 2 i \varphi},
\label{BRDEC21}
\end{equation}
showing that the vector and tensor modes 
excite, respectively, the two harmonics given in Eq. (\ref{BRDEC21}). In Fourier space  Eqs. (\ref{BRDEC6}), (\ref{BRDEC7}) and (\ref{BRDEC8}) become
\begin{eqnarray}
 {\mathcal L}_{I}^{(\mathrm{s})}(\mu, \varphi,\vec{k},\tau) &=& \partial_{\tau}\Delta^{(\mathrm{s})}_{I} + (i k \mu + \epsilon') \Delta^{(\mathrm{s})}_{I}  + i k\mu \phi  - \partial_{\tau} \psi,
\label{BRDEC6a}\\
 {\mathcal L}_{I}^{(\mathrm{v})}(\mu,\varphi,\vec{k},\tau) &=& \partial_{\tau} \Delta^{(\mathrm{v})}_{I} +  (i k \mu + \epsilon')\Delta^{(\mathrm{v})}_{I}
 \nonumber\\
 &+& \sqrt{\frac{2\pi}{3}} i\, \mu\,\biggl[ \partial_{\tau} W_{L}(\vec{k},\tau) \, Y_{1}^{-1}(\vartheta, \varphi) - \partial_{\tau} W_{R}(\vec{k},\tau) Y_{1}^{1}(\vartheta, \varphi)\biggr],
\label{BRDEC7a}\\
 {\mathcal L}_{I}^{(\mathrm{t})}(\mu,\varphi,\vec{k},\tau) &=& \partial_{\tau}\Delta^{(\mathrm{t})}_{I} + (i k \mu + \epsilon') \Delta^{(\mathrm{t})}_{I} 
 \nonumber\\
&-& \sqrt{\frac{2\pi}{15}} \biggl[ \partial_{\tau}h_{R}(\vec{k},\tau) 
Y_{2}^{2}(\vartheta, \varphi) +  \partial_{\tau} h_{L}(\vec{k},\tau) Y_{2}^{-2}(\vartheta, \varphi)\biggr].
\label{BRDEC8a}
\end{eqnarray}
Similarly Eq. (\ref{BRDEC11a}) becomes, in Fourier space,   
\begin{equation}
{\mathcal L}^{(\mathrm{y})}_{X}(\mu,\varphi,\vec{k},\tau) = \partial_{\tau} \Delta^{(\mathrm{y})}_{X}+ ( i k \mu +\epsilon') \Delta^{(\mathrm{y})}_{X}. 
\label{BRDEC11b}
\end{equation}
The explicit form of the transport equations for the scalar and tensor modes of the  geometry will be scrutinized in the remaining part of this 
section. The vectors as well as the circular polarization will be neglected since they play no role in the present considerations.
\renewcommand{\theequation}{C.\arabic{equation}}
\setcounter{equation}{0}
\section{Stochastic Faraday Rotation}
\label{APPC}
\subsection{Solutions of the evolution equations}
The solutions of the evolution equations discussed in sections \ref{sec4} and \ref{sec5} can be obtained with the techniques already 
discussed in \cite{RC1}.  For equal times (but for different Fourier modes) the fluctuations of the brightness perturbations are random with the power spectrum determined by the  
(nearly scale-invariant) spectrum of (Gaussian) curvature perturbations \cite{WMAP9}. Thus, in the absence of Faraday mixing, 
$\delta^{(\mathrm{s})}_{\pm}$ obeys then a deterministic evolution in time while the spatial fluctuations of the polarization are randomly distributed and fixed by the correlation properties of the adiabatic curvature perturbations. Conversely since $X_{F}(\tau)$ is now treated as a stochastic process, the evolution equation for the polarization becomes a stochastic differential equation \cite{stochde1,stochde2,stochde3} in time and its formal solution is obtainable by iteration: 
\begin{equation}
\delta_{\pm}(\vec{k},\tau) = \sum_{n=0}^{\infty} \delta_{\pm}^{(n)}(\vec{k},\tau), \qquad \delta_{\pm}^{(0)}(\vec{k},\tau)= \delta_{P}(\vec{k},\tau).
\label{un4}
\end{equation}
For the sake of simplicity the index referring to the scalar nature of the brightness perturbations has been dropped from $\delta_{\pm}$ but has been 
kept in the source term. We shall just restore the superscript at the very end.

Neglecting the corrections ${\mathcal O}(f_{e}^2)$, Eqs. (\ref{un2a}) and (\ref{un4}) imply the following recurrence relations: 
\begin{eqnarray}
\delta_{P}(\vec{k},\tau) &=& \frac{3 }{4}(1 - \mu^2) \int_{0}^{\tau} \, e^{- i k\mu (\tau -\tau_{1})} \,  {\mathcal K}(\tau_{1})\, S^{(\mathrm{s})}_{P}(\vec{k}, \tau_{1}),
\label{un5a}\\
\delta_{\pm}^{(n+1)}(\vec{k},\tau) &=&  \pm 2\, i \, \int_{0}^{\tau} \, 
e^{- i k \mu (\tau -\tau_{1})}\, {\mathcal K}(\tau_{1}) \, X_{F}(\tau_1) \, \delta_{\pm}^{(n)}(\vec{k},\tau_{1}).
\label{un5b}
\end{eqnarray}
The differential optical depth directly enters the visibility function giving the probability that a photon is emitted between $\tau$ and $\tau + d\tau$:
\begin{equation}
{\mathcal K}(\tau_{1})=  \epsilon'(\tau_{1})\, e^{ - \epsilon(\tau_{1}, \tau)}, \qquad \epsilon(\tau_{1},\tau) = 
\int_{\tau_{1}}^{\tau} x_{e} \,\tilde{n}_{e}\, \sigma_{e\, \gamma}\, \frac{a(\tau')}{a_{0}}.
\label{un6}
\end{equation}
The full solution of Eq. (\ref{un2a}) is formally expressible as: 
\begin{eqnarray}
\delta^{(\mathrm{s})}_{\pm}(\vec{k},\tau) &=& \frac{3}{4} (1 - \mu^2) \int_{0}^{\tau}  \,e^{- i k\mu (\tau -\tau_{1})}\,  {\mathcal K}(\tau_{1})\, S^{(\mathrm{s})}_{P}(\vec{k},\tau_{1})  \, 
{\mathcal A}_{\pm}(\tau, \tau_1) \,d\tau_{1},
\nonumber\\
{\mathcal A}_{\pm}(\tau, \tau_1) &=& e^{\mp 2 \, i\, \int_{\tau_{1}}^{\tau} X_{F}(\tau') \, d\tau'}.
\label{SS}
\end{eqnarray}
The same formal solutions discussed in the scalar case can also be obtained in the tensor case. In particular we shall have that Eq. (\ref{Boltz84}) 
can be solved  by iteration as 
\begin{equation}
\overline{\delta}^{(\mathrm{t})}_{\pm}(\vec{k},\tau) = - \int_{0}^{\tau}  \,e^{- i k\mu (\tau -\tau_{1})}\,  {\mathcal K}(\tau_{1})\, S^{(\mathrm{t})}_{P}(\vec{k},\tau_{1})  \, 
{\mathcal A}_{\pm}(\tau, \tau_1) \,d\tau_{1},
\label{TT}
\end{equation}
which is formally similar to Eq. (\ref{SS}) even if the source terms are different. In Eqs. (\ref{SS}) and (\ref{TT}) the visibility 
function enters the integrand. For analytic estimates the visibility function has the approximate shape of a double Gaussian whose first peak arises around last scattering (i.e. for $\tau \simeq \tau_{r}$) while the second (smaller) peak occurs for the reionization epoch. The way the visibility function can be 
analytically approximated has been the subject of different studies \cite{pav1,pav2,pav3} that are relevant for a refinement of the 
present discussion.  However, as we shall argue,  the scaling properties 
of the nonlinear combinations of the polarization observables shall not be crucially affected by the details of the visibility 
function. The finite thickness of the last scattering surface does not affect the ratios between the different combinations of polarization power spectra discussed here so that the limit of sudden recombination can be safely adopted; in this limit the first and more pronounced Gaussian profile tends to a Dirac delta function.

\subsection{Cumulant expansion}
The randomness implies that $X_{F}(\tau)$ is not a deterministic variable but rather a stochastic process which is stationary insofar as the autocorrelation function $\Gamma(\tau_{1},\tau_{2}) = \langle X_{F}(\tau_{1}) X_{F}(\tau_{2}) \rangle$ only depends on time differences,  i.e. $\Gamma(\tau_{1},\tau_{2}) = \Gamma(|\tau_{1}- \tau_{2}|)$; furthermore we shall also assume that the process has zero mean, even if this 
is not strictly necessary for the consistency of the whole approach. 
If  $\tau_{b}$ defines the time-scale of variation of the brightness perturbations of the polarization observables, the physical 
situation investigated here corresponds to $\tau_{b} \gg \tau_{c}$ where $\tau_{c}$ is the correlation time-scale of $X_{F}$.
In the simplest case of a Gaussian-correlated process the autocorrelation function $\Gamma(\tau_{1} - \tau_{2}) = F(\tau_1) \tau_{c} \delta(\tau_{1} - \tau_{2})$.  
If the time scale of spatial variation of the rate is comparable with the time scale of spatial variation of the gravitational fluctuations, 
$X_{F}$ can be considered only time dependent (i.e. a stochastic process). In the opposite situation the Faraday rate must be considered fully inhomogeneous (i.e. a stochastic field).

The statistical properties of  ${\mathcal A}_{\pm}$ follow directly from the correlation properties of $X_{F}(\tau)$. If, for instance,  
$X_{F}(\tau)$ obeys a stationary and Gaussian process, for any set of $n$ Faraday rates (characterized by different conformal times) the correlator  $\biggl\langle X_{F} (\tau_{1})\, X_{F}(\tau_{2})\,  .\,.\,.\, X_{F}(\tau_{n}) \biggr\rangle $ vanishes if $n$ is odd; if $n$ is even the same correlator equals:
\begin{equation}
\sum_{\mathrm{pairings}}  \biggl\langle X_{F} (\tau_{1})\, X_{F}(\tau_{2})\biggl\rangle\,\biggl\langle X_{F} (\tau_{3})\, X_{F}(\tau_{4})\biggl\rangle
  .\,.\,.\,\biggl\langle X_{F}(\tau_{n-1})  X_{F}(\tau_{n})\biggr\rangle,
  \label{gaus}
\end{equation}
where the sum is performed over all the $(n-1)!$ pairings. In the Gaussian case, the evaluation of the averages can be performed by first 
doing the standard moment expansion and then resumming the obtained result. As an example, from the explicit expression of ${\mathcal A}_{\pm}$, it follows
that 
\begin{equation}
\langle {\mathcal A}_{\pm}(\tau, \tau_{r})\, {\mathcal A}_{\pm}(\tau, \tau_{r})\rangle = \biggl\langle e^{ \pm 4 i \int_{\tau_{r}}^{\tau} X_{F}(\tau')\, d\tau^{\prime}} \biggr\rangle = \sum_{n=0}^{\infty} \frac{(- 2 \omega_{F})^{n}}{n\,!},
\label{un7a}
\end{equation}
where $\omega_{F}$ is given by
\begin{equation}
 \omega_{F} = 4 \int_{\tau_{r}}^{\tau} \, d\tau_{1} \,  \int_{\tau_{r}}^{\tau} \, d\tau_{2} \langle X_{F}(\tau_{1}) \, X_{F}(\tau_{2}) \rangle.
 \label{cor1}
 \end{equation}
It follows from Eq. (\ref{cor1}) that even if $X_{F} \leq 1$, $\omega_{F}$ is not bound to be smaller than $1$.

The result  of Eq. (\ref{un7a}) holds in an approximate sense when the stationary process is only approximately Markovian, 
While the standard moment expansion can be formally adopted in specific cases (like the Gaussian one) it cannot be used to provide successive approximations. The reason is that any finite number of terms constitutes a bad representation of the function defined by the whole series. This difficulty is overcome with the use of the  cumulants that are certain combinations of the moments. Dropping the functions and keeping 
only their corresponding arguments we have that the relations between the ordinary moments and the cumulants (denoted by $\langle\langle ... \rangle\rangle$) is
\begin{eqnarray}
\langle X_{F}(\tau_{1})\rangle &=& \langle\langle  X_{F}(\tau_{1})\rangle\rangle, 
 \nonumber\\
\langle X_{F}(\tau_{1})\, X_{F}(\tau_{2})\rangle &=& \langle\langle X_{F}(\tau_{1})\rangle\rangle\langle\langle X_{F}(\tau_{2})\rangle\rangle 
 + \langle\langle X_{F}(\tau_{1})\, X_{F}(\tau_{2})\rangle\rangle,
 \nonumber\\
\langle  X_{F}(\tau_{1}) \, X_{F}(\tau_{2}) \, X_{F}(\tau_{3}) \rangle &=&  \langle\langle  X_{F}(\tau_{1})\rangle\rangle  \langle\langle  X_{F}(\tau_{2})\rangle\rangle\langle\langle  X_{F}(\tau_{3}) \rangle\rangle  
\nonumber\\
&+&\langle\langle  X_{F}(\tau_{1})\,  X_{F}(\tau_{2})\rangle\rangle \langle\langle  X_{F}(\tau_{3}) \rangle\rangle  
\nonumber\\
&+&\langle\langle  X_{F}(\tau_{3})\,  X_{F}(\tau_{1})\rangle\rangle \langle\langle  X_{F}(\tau_{2}) \rangle\rangle 
\nonumber\\
&+&  \langle\langle  X_{F}(\tau_{2})\,  X_{F}(\tau_{3})\rangle\rangle \langle\langle  X_{F}(\tau_{1}) \rangle\rangle 
\nonumber\\
&+&
\langle\langle  X_{F}(\tau_{1})\, X_{F}(\tau_{2})\, X_{F}(\tau_{3})\rangle\rangle,
 \label{cor1a}
 \end{eqnarray}
 and so on and so forth for the other moments of the cluster expansion. Substituting the naive moment expansion with the cumulant expansion we have that the average of Eq. (\ref{un7a}) is given by
\begin{equation}
\langle {\mathcal A}_{\pm}(\tau, \tau_{r})\, {\mathcal A}_{\pm}(\tau, \tau_{r})\rangle = \exp{\biggl[ \sum_{m=1}^{\infty}   \frac{(\pm 4 i)^{m}}{m!} \int_{\tau_{r}}^{\tau} d\tau_{m}\, \, \biggl\langle\biggl\langle X_{F} (\tau_{1})\, X_{F}(\tau_{2})\,  .\,.\,.\, X_{F}(\tau_{m}) \biggr\rangle \biggr\rangle\biggr]}.
\label{cor4}
\end{equation}
 As firstly suggested by Van Kampen (see Ref. \cite{stochde1,stochde3})
in the approximately Markovian case the averages of certain stochastic processes will be given by an exponential whose exponent is a series of successive cumulants of $X_{F}$.  All the cumulants beyond the second are zero in the case of an exactly Gaussian process and 
the result reported in Eq. (\ref{un7a}) is recovered. Since each integrand in (\ref{cor4}) virtually vanishes unless $\tau_{1}$, $\tau_{2}$,..., $\tau_{m}$ are close
together, the only contribution to the integral comes from a tube of diameter of order $\tau_{c}$ along the diagonal in the $m$-dimensional integration space. 
More generally, the $m$-th cumulant vanishes as soon as the sequence of times $\tau_{1}$, $\tau_{2}$,...,$\tau_{m}$ contains a gap large compared to $\tau_{c}$.
This is the reason why, in a nutshell, a cumulant is also rather practical in our case. 
 
If the stochastic process is not homogeneous the iterative solution of Eqs. (\ref{un4}) and (\ref{un5a})--(\ref{un5b}) can be written as:
\begin{eqnarray}
&&\partial_{\tau} \delta_{\pm}^{(0)} + ( i k \mu + \epsilon') \delta_{\pm}^{(0)} = \frac{3}{4}(1 -\mu^2) \epsilon' S_{P}(\vec{k},\tau),
\label{nh1}\\
&& \partial_{\tau} \delta_{\pm}^{(1)} + ( i k \mu + \epsilon') \delta_{\pm}^{(1)} = \mp i \, b_{F}(\overline{\nu},\tau) \, \int d^{3} p \, \delta_{P}(\vec{k} + \vec{p}, \tau) \, n^{i} \, B_{i}(\vec{p},\tau),
\label{nh2}\\
&&  \partial_{\tau} \delta_{\pm}^{(2)} + ( i k \mu + \epsilon') \delta_{\pm}^{(2)} = \mp i \, b_{F}(\overline{\nu},\tau) \, \int d^{3} p^{\prime} \, \delta_{P}(\vec{k} + \vec{p}^{\,\prime}, \tau) \, n^{i} \, B_{i}(\vec{p},\tau) \, n^{j}\, B_{j}(\vec{p}^{\,\prime}, \tau),
\label{nh3}
\end{eqnarray}
To compute the averages we must therefore specify the correlation properties of the Faraday rate. Even if the spatial dependence may reside 
in all the terms contributing to the Faraday rate, it is reasonable to presume that the leading effect may come from the magnetic field whose 
correlation function will then be parametrized as:
\begin{equation}
\langle B_{i}(\vec{q},\,\tau_{1}) \, B_{j}(\vec{p},\,\tau_{2}) \rangle = \frac{2 \pi^2}{p^3} P_{ij}(\hat{p}) \, \overline{P}_{B}(p) \, \Gamma(|\tau_{1} - \tau_{2}|) \, \delta^{(3)}(\vec{q} + \vec{p}),
\label{CI1}
\end{equation}
where $\Gamma(|\tau_{1} - \tau_{2}|) = \tau_{c}\, \delta(\tau_{1} - \tau_{2})$ in the delta-correlated case. In the same approximation exploited before and using Eq. (\ref{CI1}),  $\omega_{F}$ becomes now 
\begin{equation}
\omega_{F} = \frac{8 \overline{b}_{F}^2}{3 \overline{\nu}^4} \, \int \frac{d p}{p} \overline{P}_{B}(p) \, \int_{\tau_{r}}^{\tau} d\tau_{1} \int_{\tau_{r}}^{\tau} d\tau_{2} \frac{\Gamma(|\tau_{1} - \tau_{2}|)}{a^2(\tau_{1}) 
a^2(\tau_{2})},
\label{CI2}
\end{equation}
where the constant $\overline{b}_{F}= b_{F}(\overline{\nu},\tau)a^2(\tau) \overline{\nu}^2$ has been introduced in order to draw special attention 
to the frequency scaling which is the most relevant aspect of Eq. (\ref{CI2}), at least in the present approach. 

\end{appendix}

\end{document}